\documentclass[aps,pre,reprint,groupedaddress]{revtex4-2}
\usepackage[colorlinks=true,citecolor=blue]{hyperref}

\usepackage{amsmath}
\usepackage{amssymb}
\usepackage{amsthm}
\usepackage{bbm}
\usepackage{verbatim}
\usepackage{graphicx}
\usepackage{mathtools}
\usepackage{color}

\newtheorem{theorem}{Theorem}

\newcommand{\sA}{\mathsf{A}}
\newcommand{\sB}{\mathsf{B}}

\newcommand{\sD}{\mathsf{D}}
\newcommand{\sE}{\mathsf{E}}
\newcommand{\sF}{\mathsf{F}}
\newcommand{\sG}{\mathsf{G}}
\newcommand{\sH}{\mathsf{H}}

\newcommand{\sK}{\mathsf{K}}
\newcommand{\sL}{\mathsf{L}}
\newcommand{\sM}{\mathsf{M}}
\newcommand{\sN}{\mathsf{N}}
\newcommand{\sP}{\mathsf{P}}
\newcommand{\sQ}{\mathsf{Q}}
\newcommand{\sR}{\mathsf{R}}
\newcommand{\sS}{\mathsf{S}}
\newcommand{\sT}{\mathsf{T}}
\newcommand{\sU}{\mathsf{U}}
\newcommand{\sV}{\mathsf{V}}
\newcommand{\sW}{\mathsf{W}}
\newcommand{\sX}{\mathsf{X}}
\newcommand{\sY}{\mathsf{Y}}
\newcommand{\sZ}{\mathsf{Z}}

\newcommand{\mB}{\mathcal{B}}

\newcommand{\mL}{\mathcal{L}}

\newcommand{\mR}{\mathcal{R}}
\newcommand{\mT}{\mathcal{T}}
\newcommand{\mU}{\mathcal{U}}

\newcommand{\kp}{\kappa}

\newcommand{\lb}{\lambda}
\newcommand{\oa}{\omega}
\newcommand{\sg}{\sigma}

\newcommand{\rv}{\mathrm{v}}
\newcommand{\rx}{\mathrm{x}}

\newcommand{\nrm}[1]{\lVert #1 \rVert}
\newcommand{\abs}[1]{| #1 |}

\renewcommand{\Re}[1]{\mathrm{Re}[#1]}

\newcommand{\bC}{\mathbb{C}}
\newcommand{\bR}{\mathbb{R}}
\newcommand{\bM}{\mathbb{M}}
\newcommand{\bN}{\mathbb{N}}

\newcommand{\bCn}{\mathbb{C}^{N\times N}}

\newcommand{\tint}{\int_0^t \! dt' \;}
\newcommand{\fint}{\int_0^\infty \! dt \;}
\newcommand{\lint}{\int_0^1 \! d\lambda \;}

\newcommand{\trans}{\mathsf{T}}

\begin{document}

\title{Dynamics of micro and nanoscale systems in the weak-memory regime:\\ 
A mathematical framework beyond the Markov approximation}


\author{Kay Brandner}
%\thanks{}
\affiliation{
School of Physics and Astronomy, 
University of Nottingham, 
Nottingham, 
NG7 2RD, United Kingdom
}
\affiliation{
Centre for the Mathematics and Theoretical Physics of Quantum Non-Equilibrium Systems,
University of Nottingham, 
Nottingham, 
NG7 2RD, United Kingdom
}


\date{\today}

\begin{abstract}
The visible dynamics of small-scale systems are strongly affected by unobservable degrees of freedom, which can belong either to external environments or internal subsystems and almost inevitably induce memory effects. 
Formally, such inaccessible degrees of freedom can be systematically eliminated from essentially any microscopic model through projection operator techniques, which result in non-local time evolution equations.
This article investigates how and under what conditions locality in time can be rigorously restored beyond the standard Markov approximation, which generally requires the characteristic time scales of accessible and inaccessible degrees of freedom to be sharply separated. 
Specifically, we consider non-local time evolution equations that are autonomous and linear in the variables of interest. 
For this class of models, we prove a mathematical theorem that establishes a well-defined weak-memory regime, where faithful local approximations exists, even if the relevant time scales are of comparable order of magnitude.
The generators of these local approximations, which become exact in the long-time limit, are time independent and can be determined to arbitrary accuracy through a convergent perturbation theory in the memory strength, where the Markov generator is recovered in first order.  
For illustration, we work out three simple, yet instructive, examples covering coarse grained Markov jump networks, semi-Markov jump processes and generalized Langevin equations.  
\end{abstract}

%\keywords{}

\maketitle

\section{Introduction}\label{sec:I}
\subsection{Physical Setting}\label{sec:Ips}
Almost any physical system is open in that its accessible degrees of freedom are coupled to inaccessible ones, which cannot be directly observed. 
Such inaccessible degrees of freedom induce fluctuations and memory effects, both of which are particularly pronounced on micro and nanoscales.
The erratic motion of Brownian particles, for instance, is caused by collisions with invisible fluid molecules in their surroundings.  
By pushing aside the fluid in their vicinity, the particles alter the momentum and density of the molecules with which they collide at later times, an effect is known as hydrodynamic memory \cite{berg-sorensen2005}. 
As a result, an observer can only predict the trajectories of the particles, or more generally the accessible degrees of freedom, with a certain probability and how these trajectories evolve depends on their history.
In formal terms, open systems are described by probabilistic time evolution equations that are non-local in time \cite{fick1990,grabert1992,zwanzig2001,vrugt2020}.
 
Yet, our understanding of open systems is to a large extent based on the premise that the immediate future of the accessible degrees of freedom depends only on their present state \cite{vankampen2007,risken1996,breuer2002,strasberg2022,
shiraishi2023}.
This assumption leads to time-local equations of motion, which provide very good approximations of their non-local counterparts if two physical conditions are met \cite{fick1990,grabert1992,zwanzig2001,vrugt2020}. 
First, there must be a clear separation of time scales that makes it possible to distinguish slow and fast evolving degrees of freedom. 
Second, all inaccessible degrees of freedom must be fast such that they relax to a generic state before causing any significant back action on the accessible ones. 
These conditions are often very well satisfied. 
The momenta of Brownian particles, for example, usually change much slower than those of the much lighter fluid molecules, which, on the time scale of the observer, return instantly to thermal equilibrium after being displaced by the particles.
More generally, the environments of classical small-scale systems, which, besides Brownian particles, include chemical reaction networks, bio-molecular machines and nanoscale electronic devices in the Coulomb-blockade regime, are typically fast-relaxing thermal reservoirs, which do not induce any significant memory effects \cite{seifert2012,benenti2017}. 

Still, there is a large variety of physical settings where memory effects play an important role \cite{trushechkin2022,klippenstein2021,lapolla2019,breuer2016,
hofling2013} and may even be directly observed with modern experimental techniques \cite{mehl2012,fanchini2014,groblacher2015,garcia-perez2020,chen2022,narinder2018,ginot2022,jeney2008, franosch2011}. 
For Brownian motion, this situation occurs, for example, when the hydrodynamic memory is amplified by a viscoelastic medium \cite{narinder2018,ginot2022}, by constraining geometries or strong confinement potentials \cite{jeney2008,franosch2011}. 
In general, memory effects become important whenever there is no strong separation of time scales between accessible and inaccessible degrees of freedom, where the latter can either belong to an external environment or an internal subunit of the open system. 
The thermodynamic properties of mesoscopic reservoirs, for instance, may change rapidly in response to the exchange of energy and matter, which can lead to a retarded back action on the system of interest \cite{gallego-marcos2014,grenier2016,krinner2017,amato2020,pekola2021,
riera-campeny2021,strasberg2021,yuan2022,schaller2014,
brange2018,moreira2023}. 
In a similar way, memory effects can be induced by hidden subsystems such as the internal configurations of complex bio-molecules \cite{hummer2015,martinez2019,hartich2021,ayaz2021,ertel2022,hartich2023,zhao2024}.

In principle, any system with finite memory has to be described with non-local time evolution equations. 
In practice, however, one usually seeks an effective description in terms of local equations, which tend to be physically more transparent and are easier to handle both analytically and numerically. 
To this end, a large variety of techniques has been developed over the last decades, some of which we will consider in more detail in Sec.~\ref{sec:P}, see for instance Refs.~\cite{nestmann2021a,nestmann2021,bruch2021,contreras-pulido2012,karlewski2014,
schaller2008,schaller2009,majenz2013,farina2019,mozgunov2020,cattaneo2019,
nathan2020,davidovic2020,kleinherbers2020,trushechkin2021,davidovic2022}.
However, most of the existing approaches do not address questions of convergence and uniqueness, rely at least partly on phenomenological arguments or still require a strong separation of time scales;  
that is, one has to assume that the characteristic time scales of accessible and inaccessible degrees of freedom differ by several orders of magnitude. 
As a result, it is not entirely clear under which conditions a given non-local time evolution equation actually admits a faithful local approximation, whether this approximation is unique, how it can be systematically constructed and how much the resulting effective dynamics deviates from the actual time evolution of the system. 
Making progress in these directions is the main objective of this article. 
As we will show, the above questions can be answered rigorously for a quite general class of systems, where memory effects are significant but still subdominant. 

\subsection{Formal Approach}\label{sec:Ifa}

For the sake of clarity, it is useful to further explain our approach in slightly more technical terms. 
As our starting point, we consider a generic non-local time evolution equation of the form \cite{fick1990,grabert1992,zwanzig2001,vrugt2020}
\begin{equation}\label{eq:int:nleeq}
	\dot{x}_t = \sV x_t + \tint \sK_{t'} x_{t-t'}. 
\end{equation}
Here, the finite-dimensional vector $x_t$ describes the state of the accessible degrees of freedom at the time $t$, for instance in terms of a discrete probability distribution, the matrix $\sV$ generates the effective evolution that would be observed if the inaccessible degrees of freedom would relax infinitely fast, and the matrix function $\sK_t$, which we require to be continuous at $t=0$, accounts for memory effects. 
We take $\sV$ to be time independent and $\sK_t$ to depend only on the memory but not on the present time, which is often the case if no time dependent driving fields are applied to the system. 
Equations of the type \eqref{eq:int:nleeq} are know as liner Volterra integro-differential equations in the mathematical literature, where the general properties of their solutions, such as existence, uniqueness and stability, have been thoroughly studied \cite{lakshmikantham1995,burton2005}. 
In physics, such equations arise when unobservable or irrelevant degrees of freedom are eliminated from local time evolution equations with the linear structure  
\begin{equation}\label{eq:int:sleeq}
	\dot{X}_t = \sW X_t. 
\end{equation}
Here, the extended state vector $X_t$ describes the full model with all its degrees of freedom and the microscopic generator $\sW$ corresponds to a time independent matrix, which can be infinite-dimensional as long as the state space of the relevant degrees of freedom is discrete and finite. 
The transition from Eq.~\eqref{eq:int:sleeq} to Eq.~\eqref{eq:int:nleeq} can be achieved systematically with  projection operator techniques, where it is often assumed that the inaccessible degrees of freedom are initially in a stationary state with respect to their unperturbed dynamics \cite{fick1990,grabert1992,zwanzig2001,vrugt2020}.
Such procedures are known as coarse graining and can be applied, for instance, to Markov jump networks or microscopic models of open quantum systems
\cite{landman1977,esposito2012,bo2017,strasberg2019,seiferth2020}.

For systems with continuous degrees of freedom, like Brownian particles, the situation is more complicated, as the underlying micro-dynamics may be non-linear. 
Still, projection operator methods can be formally adapted to non-linear settings, for example by treating Poisson brackets as linear operators \cite{zwanzig2001}. 
In this way, it is still possible to derive linear time evolution equations for finite sets of observables.
Such Mori-type generalized Langevin equations take, up to an inhomogeneous noise term, which can be systematically dealt with, the form of Eq.~\eqref{eq:int:nleeq}, at least if the original model is Hamiltonian and the eliminated degrees of freedom are initially in thermal equilibrium \cite{zwanzig2001}.  
However, under more general conditions, like stochastic or non-autonomous micro-dynamics, accurate descriptions of the reduced system usually require more involved time evolution equations, which may be non-linear in the relevant observables or feature an explicitly time dependent memory kernel \cite{lapolla2019,schilling2022}. 

Leaving these more advanced settings to future research, we now return to Eq.~\eqref{eq:int:nleeq}. 
Once an effective time evolution equation of this form has been obtained, one usually seeks to simplify it through suitable approximations, which may depend on the specifics of the considered setup and often rely on comparing relevant time scales.
A sharp separation of time scales is realized if the characteristic decay time of $\sK_t$ is orders of magnitude smaller when the remaining time scales of the system. 
It is then plausible to assume that the solution $x_t$ of Eq.~\eqref{eq:int:nleeq} can be approximated by the solution $y_t$ of a local time evolution equation with the form 
\begin{equation}\label{eq:int:leeq}
	\dot{y}_t = \sL y_t,
\end{equation}
where $\sL$ is a time independent matrix or generator. 
The simplest way to achieve this reduction is to neglect the memory term entirely.
This approach is often called adiabatic approximation and leads to $\sL^0=\sV$, which is why we will refer to $\sV$ as the adiabatic generator \cite{esposito2012,bo2017,strasberg2019,seiferth2020}. 
The Markov approximation is more accurate in that memory effects are not neglected but averaged out, which yields the Markov generator \cite{hummer2015,fick1990}
\begin{equation}\label{eq:int:MarkovGen}
	\sL^1 = \sV + \fint \sK_t e^{-\sV t}. 
\end{equation}
Still, also the Markov approximation is of phenomenological type, since it does not result from a systematic expansion in some small parameter in any obvious way. 
It is therefore \emph{a priori} unclear what its actual range of validity is, or how it could, at least in principle, be improved by including higher-order corrections. 

In the following, we will develop a general mathematical framework that makes it possible to address these problems. 
To this end, we focus on systems whose memory is still short-ranged but can no longer be entirely neglected or averaged out. 
This assumption is reflected by the mathematical condition that the memory kernel $\sK_t$ decays exponentially at long times, that is, 
\begin{equation}\label{eq:int:bound}
	\nrm{\sK_t} \leq M e^{-kt} \quad \text{for any} \quad t\geq 0,
\end{equation}
where $\nrm{\cdot}$ is a matrix norm and $M$ and $k$ are positive constants. 
This condition will turn out to be sufficient to make the question whether there exists a unique time independent generator mathematically well defined. 
It also introduces two different time scales $1/k$ and $1/\sqrt{M}$, which correspond to the memory time of the system and the time scale on which the accessible degrees of freedom interact with the inaccessible ones. 
From these quantities, we can construct a dimensionless parameter $\varphi = M/k^2$, which should be small if memory effects are weak. 
In fact, we will see that the adiabatic and the Markov generator correspond to zeroth and first-order truncations, respectively, of a systematic expansion of the generator $\sL$ in this parameter. 
Moreover, this expansion can be shown to converge if 
\begin{equation}\label{eq:int:cond}
	 v<k \quad\text{and}\quad 4M < (k-v)^2,
\end{equation} 
where $v= \nrm{\sV}$ denotes the third relevant time scale of the problem, which characterizes the intrinsic dynamics of the accessible degrees of freedom.
The solution of Eq.~\eqref{eq:int:nleeq} then admits a faithful local approximation, which becomes exact at long times and is given by 
\begin{equation}\label{eq:int:LTA}
	y_t = e^{\sL t} \sD x_0. 
\end{equation}
Here, $\sD$ is time independent non-singular matrix, which, like the generator $\sL$, can be determined through a convergent perturbation theory in the memory strength $\varphi$. 
In physical terms, the conditions \eqref{eq:int:cond} require that the memory of the system decays faster than its state changes due to both intrinsic dynamics and perturbations by the inaccessible degrees of freedom. 
Notably, however, the memory time $1/k$ does not have to be separated from the remaining time scales $1/v$ and $1/\sqrt{M}$ by several orders of magnitude. 
We therefore, refer to the class of systems that satisfy the requirements \eqref{eq:int:bound} and \eqref{eq:int:cond} as systems with a weak separation of time scales, or weak memory. 

Three remarks are in order before we enter the main part of this article. 
First, we note that dynamics with memory are commonly referred to as non-Markovian. 
However, time evolution equations of the type \eqref{eq:int:nleeq} and \eqref{eq:int:leeq} do not necessarily have to describe proper non-Markovian and Markovian stochastic processes, respectively \cite{vacchini2011b}. 
Since our analysis will depend only on the algebraic structure of Eq.~\eqref{eq:int:nleeq}, we here use the less specific terms local and non-local rather than Markovian and non-Markovian. 
Second, introducing a so-called slippage matrix $\sD$ to shift the initial condition of the system is a well known technique, which has been previously used to improve the accuracy of the Markov approximation \cite{geigenmuller1983,haake1983,haake1985,suarez1992,gaspard1999,
yu2000}.
As we will see in the following, this correction is in fact crucial to minimize deviations between the exact solution of Eq.~\eqref{eq:int:nleeq} and the long-time approximation \eqref{eq:int:LTA}. 
Third, it is well established that, under some technical conditions, Eq.~\eqref{eq:int:nleeq} is formally equivalent to a local time evolution equation of the form \cite{tokuyama1975,tokuyama1976,grabert1977,grabert1978,chaturvedi1979,
shibata1980,shibata1977,chaturvedi1979,shibata1980,breuer2001}
\begin{equation}
	\dot{x}_t = \sL_t x_t.
\end{equation}
The time dependent generator $\sL_t$, which is known as time-convolutionless or TCL generator, is unique and can be calculated perturbatively \cite{shibata1977,chaturvedi1979,shibata1980,breuer2001}, though it is in general not clear under which conditions theses expansions converge. 
The framework developed in this article should be seen as complementary to the TCL method, which we will revisit in Sec.~\ref{sec:FpfTCL}.
On the one hand, our approach is less general since it relies on the conditions \eqref{eq:int:bound} and \eqref{eq:int:cond}. 
On the other hand, however, our theory makes it possible to explicitly separate the effective local time evolution of the system from residual memory effects. 
Specifically, we will show that Eq.~\eqref{eq:int:nleeq} is equivalent to a quasi-local equation of motion with the form
\begin{equation}\label{eq:int:effeq}
\dot{x}_t = \sL x_t + \sE_t x_0,
\end{equation}
where $\sL$ is a time independent, or local, generator, the memory function $\sE_t$ decays exponentially in time and $x_0$ is the initial state of the system.

\subsection{Outline}\label{sec:Iol}

This article provides a self-contained expansion of the core results reported in Ref.~\cite{brandner2024a} and delivers the detailed mathematical proofs that have been skipped there for conciseness. 
We proceed as follows. 
To build intuition for the problem at hand, we use the following section to analyze simple toy models. 
In Sec.~\ref{sec:Ftt}, we present our first main results, which are summarized in a master theorem addressing the existence and uniqueness of local generators and the accuracy of the corresponding long-time approximations. 
We then provide a detailed proof of this theorem in Sec.~\ref{sec:Fpf}, which may be skipped by readers who are less interested in the technical aspects of this work. 
In Sec.~\ref{sec:Fpt}, we show how the generators can be systematically determined through a perturbation theory in the memory strength, before briefly discussing non-perturbative methods to access some of the properties of the these generators and providing several simple extensions of our main theorem in Secs.~\ref{sec:Fnp} and \ref{sec:Fex}, respectively. 
We illustrate our theory with simple models of a molecular motor, a quantum dot in contact with a mesoscopic reservoir and memory-affected Brownian motoin in Sec.~\ref{sec:E}. 
These examples are not meant to explore the full range of implications of our theory, but rather to illustrate several of its aspects in a transparent and straightforward manner. 
Finally, we further discuss connections with the existing literature and highlight some key problems for future research in Sec.~\ref{sec:P}.

\section{A First Approach}\label{sec:A}
To gain a first understanding of the structure of the solutions of the non-local time evolution equation \eqref{eq:int:nleeq}, it is instructive to consider its scalar instance 
\begin{equation}\label{eq:FA:nleeq}
	\dot{x}_t = V x_t + \tint K_{t'} x_{t-t'}. 
\end{equation}
Here, $x_t\in\bC$ is a time dependent state variable, which takes the given value $x_0\neq 0$ at the initial time $t=0$. 
Furthermore, $V\in\bC$ is a scalar adiabatic generator and the memory kernel $K_t\in\bC$ is a function of time defined for $t\geq 0$. 
Since we are interested in the weak-memory regime, we require that $K_t$ decays exponentially at long times. 
That is, we assume that there exist strictly positive constants $M$ and $k$ such that 
\begin{equation}
	\abs{K_t}\leq M e^{-kt}.
\end{equation} 
Our aim is to explore under what conditions and in what sense the solution of Eq.~\eqref{eq:FA:nleeq} can be approximated by a solution of the local time evolution equation 
\begin{equation}\label{eq:FA:leeq}
	\dot{y}_t = L y_t,
\end{equation}
where $L\in\bC$ is a time independent scalar generator. 
We first note that the short-time solutions of the Eqs.~\eqref{eq:FA:nleeq} and \eqref{eq:FA:leeq},
\begin{align}
	x_t & = x_0 + Vx_0 t + \frac{V^2 + K_0}{2} x_0 t^2 + \mathcal{O}(t^3),\\
	y_t & = y_0 + L y_0 t + \frac{L^2}{2} y_0 t^2 +\mathcal{O}(t^3),
\end{align} 
can in general not be identical, regardless of how we choose $L$ and $y_0$. 
It is however still possible that $x_t$ and $y_t$ approach each other at long times. 
More precisely, the solution of Eq.~\eqref{eq:FA:nleeq} may take the form 
\begin{equation}
	x_t = e^{L t} A_t x_0,
\end{equation}
where the function $A_t\in\bC$, to which we refer as reduced propagator, accounts for short-time corrections. 
If $L$ can be chosen such that $A_t$ tends to a non-zero constant $D$ in the limit $t\rightarrow\infty$, we say that a local generator exists. 
The solution of Eq.~\eqref{eq:FA:leeq} for the initial condition $y_0 = D x_0$, 
\begin{equation}
	y_t = e^{L t} D x_0,
\end{equation}
can then be regarded as a long-time approximation of $x_t$ and we refer to this approximation as faithful if
\begin{equation}
	\lim_{t\rightarrow\infty} |x_t - y_t| = 0. 
\end{equation}

For the sake of concreteness, we now set $V=-v<0$
and focus on the simple memory kernel
\begin{equation}\label{eq:FA:MemKer}
	K_t = - M e^{-kt}.
\end{equation}
The solutions of Eq.~\eqref{eq:FA:nleeq}, which can be easily found by Laplace transformation, then reads  
\begin{equation}
	x_t	= \sum_{j=1}^2 \frac{k+\lambda_j}{\lambda_j-\lambda_{j+1}}e^{\lambda_j t} x_0,
\end{equation}
where $\lambda_{2+1}=\lambda_1$ and
\begin{equation}
	\lambda_j = - \frac{k+v}{2}-(-1)^j\frac{\sqrt{(k-v)^2-4M}}{2} 
\end{equation}
For a local generator to exist, the two characteristic rates $\lambda_1$ and $\lambda_2$ must have distinct real parts, which is the case if and only if 
\begin{equation}\label{eq:FA:Cond1}
	4M < (k-v)^2. 
\end{equation}
If this condition is met, $L$ can be uniquely identified with the rate $\lambda_1$, which is strictly larger than $\lambda_2$. 
The reduced propagator $A_t = e^{-L t}x_t/x_0$ then converges to the non-zero constant $D= (k+\lambda_1)/(\lambda_1-\lambda_2)$ in the limit $t\rightarrow\infty$ and the long-time approximation $y_t$ is faithful, as can be easily verified by noting that $\lambda_2<0$. 

\begin{figure}
\includegraphics[width=4.25cm]{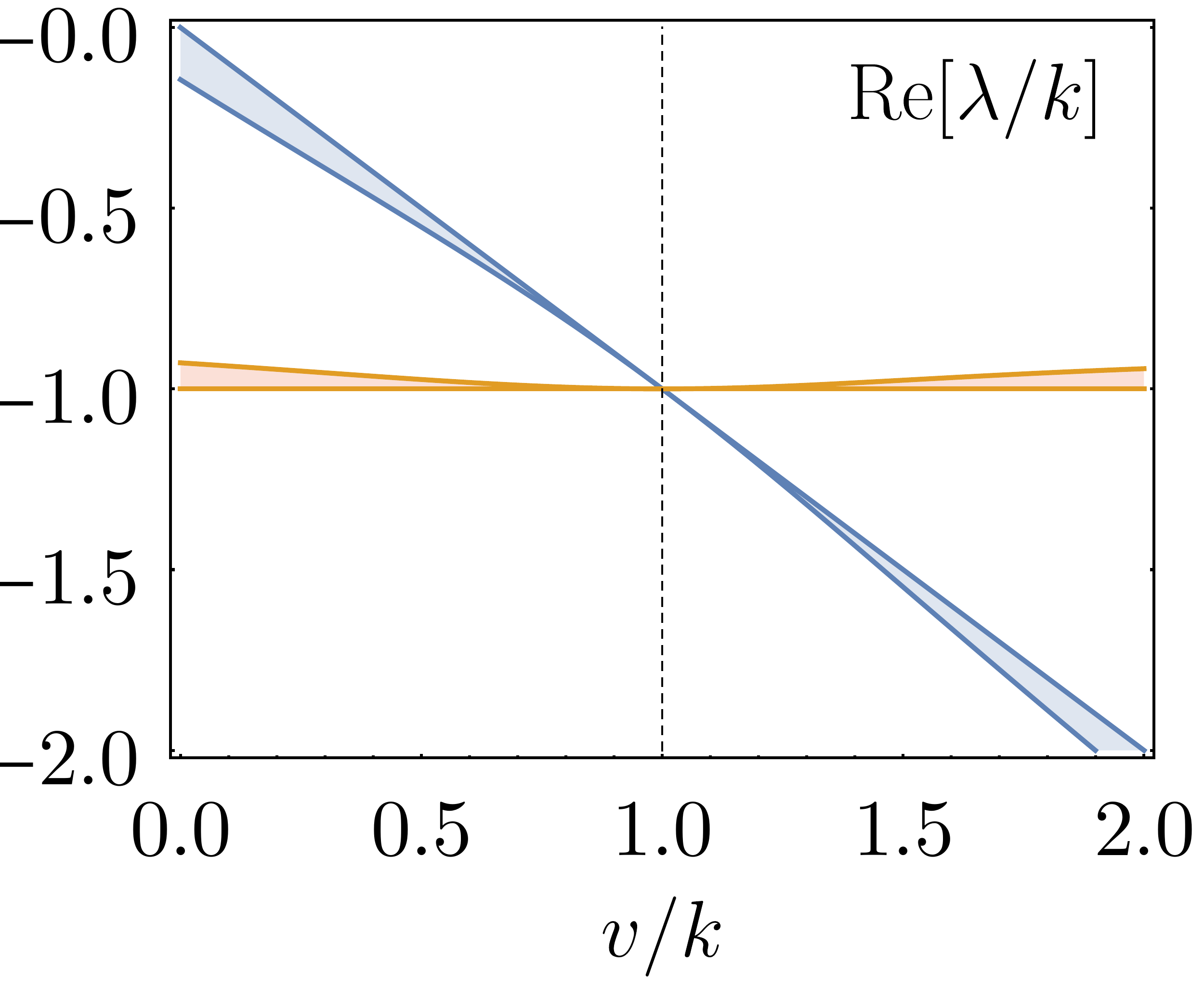}
\includegraphics[width=4.25cm]{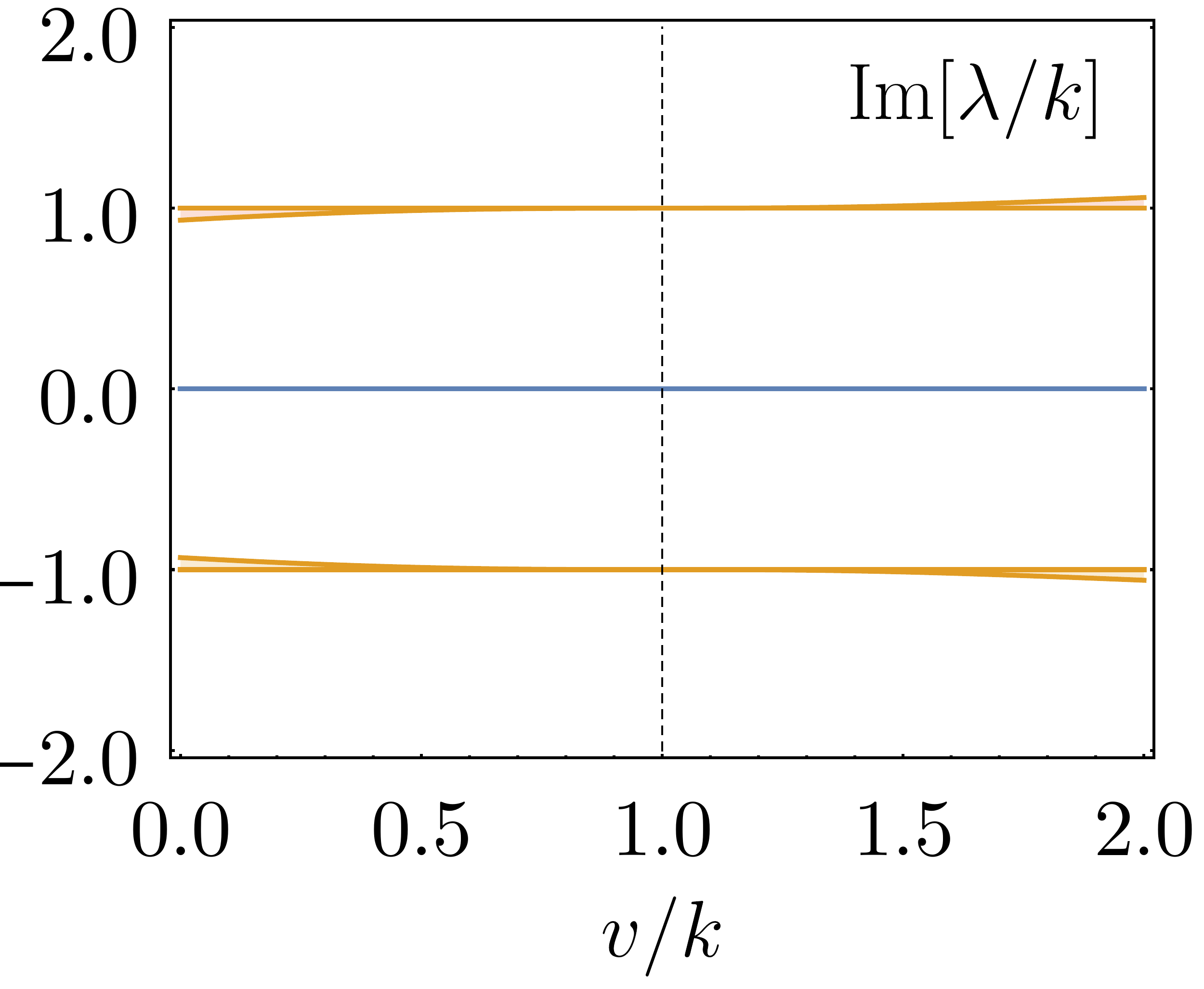}
\caption{Roots of the polynomial \eqref{eq:FA:Poly}. 
If $v\neq k$ and $M>0$ satisfy the condition \eqref{eq:FA:Cond1}, the real and the imaginary parts of the three roots of Eq.~\eqref{eq:FA:Poly} lie inside the colored areas. 
For $v\geq k$, the real part of the two complex conjugate roots, which are shown in orange, is no longer smaller than the remaining real root, which is indicated in blue, while their imaginary parts are non-zero. 
Hence, there exist at least two distinct roots with maximal real part. 
\label{Fig:FA:Rates}}
\end{figure}

The condition \eqref{eq:FA:Cond1} is satisfied if the amplitude $M$ of the memory kernel is  sufficiently small, while its overall decay rate $k$ can be either smaller or larger than the modulus $v$ of the adiabatic generator. 
In general, however, the additional condition 
\begin{equation}\label{eq:FA:Cond2}
	v<k
\end{equation}
is necessary to ensure the existence of a local generator, as the following example shows. 
Upon replacing the memory kernel \eqref{eq:FA:MemKer} with 
\begin{equation}
	K_t = - M e^{-k t}\sin[kt],
\end{equation}
the general solution of Eq.~\eqref{eq:FA:nleeq} becomes 
\begin{equation}
	x_t = \sum_{j=1}^3 
		\frac{k^2+(k+\lambda_j)^2}{(\lambda_j-\lambda_{j+1})(\lambda_j-\lambda_{j+2})}
			e^{\lambda_j t} x_0
\end{equation}
with $\lambda_{3+1}=\lambda_1$. 
Here, the characteristic rates $\lambda_j$ must be found by solving the cubic equation 
\begin{equation}\label{eq:FA:Poly}
	(v+\lambda)(k^2 + (k+\lambda)^2) + Mk = 0.
\end{equation}
A local generator exists if and only if this equation has a unique root with maximal real part. 
As the plots in Fig.~\ref{Fig:FA:Rates} show, this requirement is only met if the condition \eqref{eq:FA:Cond2} is satisfied, while the condition \eqref{eq:FA:Cond1} alone is not sufficient to this end. 
If both conditions hold, the real root of Eq.~\eqref{eq:FA:Poly} can be uniquely identified with the generator $L$. 
Since the remaining complex roots have negative real parts, the long-time approximation $y_t$ is faithful. 

Our analysis so far shows that the existence of a local generator is a non-trivial property of the adiabatic generator and the memory kernel, even if the latter is required to decay exponentially at long times. 
In the two examples above, which were selected for their mathematical simplicity and illustrative character, the conditions \eqref{eq:FA:Cond1} and \eqref{eq:FA:Cond2} were sufficient to this end. 
In both cases, these conditions were also sufficient to ensure that the generator is unique and that the corresponding long-time approximation is faithful. 
As we will see in the next section, these results are indicative of a more general mathematical structure, which makes it possible to systematically characterize the solutions of the non-local time evolution equation \eqref{eq:int:nleeq}.

\section{Framework}\label{sec:F}
\subsection{Master Theorem}\label{sec:Ftt}
We now return to our main problem discussed in the introduction.  
Our aims in this section are, first, to determine under which conditions a given solution of the non-local time evolution equation \eqref{eq:int:nleeq} can be approximated by a solution the local master \eqref{eq:int:leeq} at long times, second, to provide rigorous bounds on the error induced by this approximation and, third, to develop a systematic perturbation theory for the corresponding generators.
The first two of these problems are addressed by the following theorem, which contains our first main results. 

\noindent
\begin{theorem}\label{thr:LG}
Let $x_t\in\bC^N$ be a solution of the non-local time evolution equation
\begin{equation}\label{eq:Ftt:nleeq}
	\dot{x}_t = \sV x_t + \tint \sK_{t'} x_{t-t'}
\end{equation}
for $t\geq 0$ and given $x_0$. 
Here, $\sV\in \bCn$ denotes the adiabatic generator and the memory kernel $\sK_t \in \bCn$ is a matrix function defined for $t\geq 0$. 
Assume that there exist strictly positive constants $M$ and $k$ such that
\begin{equation}\label{eq:Ftt:kernelbound}
	\nrm{\sK_t} \leq M e^{-k t} \quad \text{for any} \quad
	t\geq 0,
\end{equation}
where $\nrm{\cdot}$ is a matrix norm on $\bCn$.  
If the conditions 
\begin{equation}\label{eq:Ftt:parabounds}
	v= \nrm{\sV} <k
		\quad\text{and}\quad
		\varepsilon= \frac{4 M}{(k-v)^2} < 1
\end{equation}
are satisfied, the following results hold.\\
\noindent
\textbf{(i)}
There exists a unique pair of local left and right generators $\sL\in\bCn$ and $\sR\in\bCn$ such that 
\begin{equation}\label{eq:Ftt:locgen}
	x_t = e^{\sL t} \sA_t x_0 = \sB_t e^{\sR t} x_0
\end{equation}
and the reduced propagators $\sA_t\in\bCn$ and $\sB_t\in\bCn$ converge to the same non-singular matrix
$\sD\in\bCn$ in the limit $t\rightarrow\infty$.
These generators satisfy the intertwining relation 
\begin{equation}\label{eq:Ftt:intertw}
\sL\sD = \sD \sR. 
\end{equation}
\noindent 
\textbf{(ii)}
The long-time approximation 
\begin{equation}\label{eq:Ftt:LTA}
	y_t = e^{\sL t} \sD x_0 = \sD e^{\sR t} x_0 
\end{equation}
of $x_t$ satisfies 
\begin{equation}\label{eq:Ftt:LTAErr}
	\abs{x_t - y_t}
	 	\leq  \frac{k-\eta}{\eta-\rho} \abs{x_0} e^{-\eta t}, 
\end{equation}
where $\abs{\cdot}$ is a vector norm on $\bC^N$ that is consistent with the matrix norm $\nrm{\cdot}$ on $\bCn$ and
\begin{align}
	\label{eq:Ftt:defrho}
	\rho & = \frac{k+v-\sqrt{(k-v)^2-4M}}{2}>0,\\
	\label{eq:Ftt:defeta}
	\eta & = \frac{k+v+\sqrt{(k-v)^2-4M}}{2}>0.
\end{align}
\end{theorem}

The first part of Theorem~\ref{thr:LG} generalizes the notions of local generators and reduced propagators from scalar variables to higher dimensional state spaces and shows that the conditions \eqref{eq:Ftt:kernelbound} and \eqref{eq:Ftt:parabounds}, which define the weak-memory regime, are sufficient for the existence of these objects.  
The examples of the previous section further demonstrate that these conditions are minimal in that, once either of them is violated, it becomes possible to construct instances of Eq.~\eqref{eq:Ftt:nleeq} for which no local generators exists. 
The second part of Theorem~\ref{thr:LG} shows that, in the weak-memory regime, the solution $y_t$ of the local time evolution equation \eqref{eq:int:leeq} for the initial condition $y_0 = \sD x_0$ is a faithful long-time approximation of $x_t$ and Eq.~\eqref{eq:Ftt:LTAErr} provides an explicit bound on the error of this approximation; an illustration of these results can be found in Sec.~\ref{sec:Emj}.
It remains to determine how the generators $\sL$ and $\sR$ can be systematically constructed.
We will return to this problem in Sec.~\ref{sec:Fpt} after proving Theorem~\ref{thr:LG}.

\subsection{Proof of Theorem 1}\label{sec:Fpf}
We proceed in four steps. 
First, we show that, if the conditions \eqref{eq:Ftt:kernelbound} and \eqref{eq:Ftt:parabounds} are fulfilled, there always exists at least one pair of local generators $\sL$ and $\sR$ that is consistent with the assertions of the first part of Theorem~\ref{thr:LG} at long times.  
We then construct the reduced propagators that correspond to these generators and show that they satisfy the asymptotic boundary condition
\begin{equation}\label{eq:Fpg:AsympBC}
	\lim_{t\rightarrow\infty} \sA_t = \lim_{t\rightarrow\infty} \sB_t
		= \sD,
\end{equation}
where the slippage matrix $\sD$ is non-singular and fulfills the intertwining relation \eqref{eq:Ftt:intertw}. 
In step three, we derive the bound \eqref{eq:Ftt:LTAErr} on the error of the long-time approximation \eqref{eq:Ftt:LTA}. 
Finally, we show that the generators and reduced propagators are unique if the latter are required to satisfy the asymptotic relation \eqref{eq:Fpg:AsympBC}. 

\subsubsection{Generators}\label{sec:FpfGen}
The propagator $\sZ_t\in\bCn$ is defined so that the solution of Eq.~\eqref{eq:Ftt:nleeq} can be written as $x_t = \sZ_t x_0$ for any initial state vector $x_0$. 
This matrix function is the unique solution of the two equivalent integro-differential equations
\begin{subequations}\label{eq:Fpg:fbwrdeq}
\begin{align}
	\label{eq:Fpg:fwrdeq}
	\dot{\sZ}_t & = \sV\sZ_t + \tint \sK_{t'}\sZ_{t-t'},\\
	\label{eq:Fpg:bwrdeq}
	\dot{\sZ}_t & = \sZ_t\sV + \tint \sZ_{t-t'}\sK_{t'}.
\end{align}
\end{subequations}
with respect to the initial condition $\sZ_0 = \mathsf{1}$. 
The first of these equations follows directly from Eq.~\eqref{eq:Ftt:nleeq}. 
The second one is known as adjoint equation and can be derived by formally solving Eq.~\eqref{eq:Fpg:fwrdeq} in Laplace space or by iteration of the corresponding integral equation \cite{lakshmikantham1995,burton2005}. 

We now assume that the propagator can be approximated as $\sZ_t = e^{\sL t}\sD = \sD e^{\sR t}$ for sufficiently large $t$. 
Inserting the first ansatz into Eq.~\eqref{eq:Fpg:bwrdeq} and the second one into Eq.~\eqref{eq:Fpg:fwrdeq} and taking the limit $t\rightarrow\infty$ yields the matrix equations \cite{geigenmuller1983}
\begin{subequations}\label{eq:Fpg:genAsmp}
\begin{align}
	\label{eq:Fpg:LAsmp}
	\sL & = \sV + \fint \sK_t e^{-\sL t},\\
	\label{eq:Fpg:RAsmp}
	\sR & = \sV + \fint e^{-\sR t} \sK_t,
\end{align}
\end{subequations}
for $\sL$ and $\sR$, where we have assumed that $\sD$ is non-singular and that the intertwining relation \eqref{eq:Ftt:intertw} holds. 
Banach's fixed point theorem now makes it possible to show that each of these equations has at least one solution \cite{agarwal2018}. 
To this end, we define the set  
\begin{equation}\label{eq:Fpg:defBrho}
	B_\rho = \bigl\{\sX\in\bCn : \nrm{\sX}\leq\rho\bigr\},
\end{equation}
where the parameter $\rho$ was introduced in Eq.~\eqref{eq:Ftt:defrho}, and two maps $T$ and $U$ so that the Eqs.~\eqref{eq:Fpg:LAsmp} and \eqref{eq:Fpg:RAsmp} can be formally written as fixed-point equations, $\sL = T\sL$ and $\sR = U\sR$. 
Next, we observe that, for $\sX \in B_\rho$, 
\begin{align}\label{eq:Fpg:closure}
	\nrm{T\sX} 	& = \nrm{\sV + \fint \sK_t e^{-\sX t}}\\
				&\leq v + M \fint e^{-(k-\nrm{\sX})t}\nonumber\\
			  	&\leq v + \frac{M}{k-\rho} = \rho, \nonumber
\end{align}
where we have used that $\nrm{e^{-\sX t}}\leq e^{\nrm{\sX}t}$ and $\nrm{\sX}\leq \rho$. 
Hence, $T$ maps the set $B_\rho$ into itself. 
Furthermore, for any $\sX, \sY\in B_\rho$, we have 
\begin{align}\label{eq:Fpg:contract1}
	\nrm{T\sX-T\sY} &=\nrm{\fint \sK_t(e^{-\sX t}-e^{-\sY t})}\\
	&\leq M \fint \nrm{e^{-\sX t}-e^{-\sY t}} e^{-k t}.\nonumber
\end{align}
The difference of the two matrix exponentials can be further bounded with the help of the identity 
\begin{align}\label{eq:Fpg:expdiff}
	& e^{\sX}-e^{\sY}  = \lint \partial_\lambda
		e^{\lambda\sX  + (1-\lambda)\sY}\\
	& = \lint \! \int_0^1 \! d\kappa \;
	    e^{[\lambda\sX  + (1-\lambda)\sY]\kappa}(\sX-\sY)
	    e^{[\lambda\sX  + (1-\lambda)\sY](1-\kappa)},\nonumber
\end{align}
which holds for any $\sX, \sY \in \bCn$ and implies 
\begin{equation}\label{eq:Fpg:expbnd}
	\nrm{e^{\sX}-e^{\sY}}\leq \nrm{\sX-\sY}
		\lint e^{\lambda\nrm{\sX}+(1-\lambda)\nrm{\sY}}.
\end{equation}
Since $\nrm{\sX},\nrm{\sY}\leq \rho$ in Eq.~\eqref{eq:Fpg:contract1}, it follows that 
\begin{align}\label{eq:Fpg:contract2}
\nrm{T\sX-T\sY} 
	& \leq M \nrm{\sX-\sY} \fint t e^{-(k-\rho)t}\\
	&= q_\varepsilon  \nrm{\sX-\sY},\nonumber 
\end{align}
where 
\begin{equation}\label{eq:Fpg:ContFact}
	q_\varepsilon = \frac{M}{(k-\rho)^2} = \frac{\varepsilon}{(1+\sqrt{1-\varepsilon})^2}
\end{equation}
and $\varepsilon$ was defined in Eq.~\eqref{eq:Ftt:parabounds}. 
Thus, since $\varepsilon<1$ by assumption, we have $q_\varepsilon<1$, which shows that $T$ is a contraction on $B_\rho$. 
Upon observing that $B_\rho$ becomes a complete metric space when equipped with the metric $d(\sX,\sY) = \nrm{\sX- \sY}$, we can therefore conclude that the map $T$ has exactly one fixed point in $B_\rho$. 
The same result holds for $U$, as can be confirmed by repeating the steps above.
That is, there are exactly one $\sL\in B_\rho$ and exactly one $\sR\in B_\rho$ that satisfy $\sL=T\sL$ and $\sR=U\sR$. 
Hence, each of the Eqs.~\eqref{eq:Fpg:genAsmp} has exactly one solution in $B_\rho$.   

From here onward, we identify these solutions with the generators of Theorem~\ref{thr:LG}. 
Once these generators are found, the intertwining relation \eqref{eq:Ftt:intertw} provides a linear equation for the slippage matrix $\sD$, whose solution is generally not unique. 
To show that there exists at least one non-singular solution, we make the ansatz 
\begin{equation}\label{eq:Fpg:DDef}
	\sD = \bigl[\mathsf{1} + \int_0^\infty \! dt \int_0^\infty dt' \;
		 e^{-\sR t}\sK_{t+t'} e^{-\sL t'}\bigr]^{-1}. 
\end{equation}
The conditions \eqref{eq:Ftt:kernelbound} and $\nrm{\sL},\nrm{\sR}\leq\rho$ ensure that the improper integrals are well defined and further imply that 
\begin{align}
	\nrm{\sD^{-1}-\mathsf{1}} & 
		\leq M \int_0^\infty \! dt \int_0^\infty dt' \; 
			e^{-(k-\rho)(t+t')} = q_\varepsilon.
\end{align}
Since $q_\varepsilon<1$, it follows that $\nrm{\sD^{-1}-\mathsf{1}}<1$, which shows that the matrix $\sD^{-1}$ is non-singular.
Hence, so is $\sD$. 
To show that $\sD$ indeed satisfies the intertwining relation \eqref{eq:Ftt:intertw}, we observe that 
\begin{align}\label{eq:Fpg:intertw}
	\sR\sD^{-1} & = \sR + \fint \sK_t e^{-\sL t}\\
		&\quad +\int_0^\infty \! dt \int_0^\infty \! dt'\; 
			e^{-\sR t}(\partial_{t'}\sK_{t+t'})e^{-\sL t'}\nonumber\\
		&= \sR + \fint \sK_t e^{-\sL t} - \fint e^{-\sR t}\sK_t \nonumber\\
		&\quad + \int_0^\infty \! dt \int_0^\infty \! dt'\; 
			e^{-\sR t} \sK_{t+t'} e^{-\sL t'}\sL \nonumber\\
		&= \sD^{-1}\sL, \nonumber
\end{align}
where we have integrated by parts twice and inserted the fixed-point equations Eqs.~\eqref{eq:Fpg:genAsmp}. 
From Eq.~\eqref{eq:Fpg:intertw}, we have the identity $\sR\sD^{-1} = \sD^{-1}\sL$, which, upon multiplying with $\sD$ from both sides, implies $\sD\sR=\sL\sD$. 

\subsubsection{Reduced Propagators}\label{sec:FpfRed}
We take $\sL$ and $\sR$ to be the generators identified in the previous step and introduce the reduced propagators $\sA_t = e^{-\sL t}\sZ_t$ and $\sB_t = \sZ_t e^{-\sR t}$. 
Upon taking a time derivative and inserting the Eqs.~\eqref{eq:Fpg:bwrdeq} and \eqref{eq:Fpg:fwrdeq}, respectively, we find that these matrix functions satisfy 
\begin{subequations}\label{eq:Fpp:TE}
\begin{align}
	\label{eq:Fpp:TEA}
	\dot{\sA}_t &= \sA_t\sV - \sL\sA_t + \tint e^{-\sL t'} \sA_{t-t'}\sK_{t'},\\
	\label{eq:Fpp:TEB}
	\dot{\sB}_t &= \sV\sB_t - \sB_t\sR + \tint \sK_{t'}\sB_{t-t'} e^{-\sR t'}, 
\end{align}
\end{subequations}
where $\sA_0 = \sB_0 = \mathsf{1}$.
These integro-differential equations are equivalent to the integral equations 
\begin{subequations}\label{eq:Fpp:IntAB}
\begin{align}
	\sA_t & = \mathsf{1} - \int_0^t \! dt' \int_{t-t'}^\infty \! dt'' \; 
		e^{-\sL(t-t')}\sA_{t'}\sK_{t''}e^{\sL(t-t'-t'')},\\
	\sB_t & = \mathsf{1} - \int_0^t \! dt' \int_{t-t'}^\infty \! dt'' \; 
		e^{\sR(t-t'-t'')}\sK_{t''}\sB_{t'}e^{-\sR(t-t')},
\end{align}
\end{subequations}
as can be verified by taking a derivative with respect to $t$ and inserting the fixed-point equations \eqref{eq:Fpg:LAsmp} and \eqref{eq:Fpg:RAsmp} to eliminate improper integrals. 

We now take $\sD$ to be the matrix introduced in Eq.~\eqref{eq:Fpg:DDef} and define $\sX'_t = \sX_t -\sD$, where $\sX=\sA,\sB$. 
Upon using the Eqs.~\eqref{eq:Fpp:IntAB} and the intertwining relation \eqref{eq:Ftt:intertw}, we find that these new matrix functions satisfy 
\begin{subequations}
\begin{align}
	\label{eq:Fpp:ApIntEq}
	\sA'_t 	& = \sD \int_t^\infty\! dt'\int_{t'}^\infty\! dt''\; 
					e^{-\sR t'}\sK_{t''}e^{\sL(t'-t'')}\\
				& \quad - \int_0^t dt' \int_{t-t'}^\infty dt'' \; 
					e^{-\sL(t-t')}\sA'_{t'}\sK_{t''}e^{\sL(t-t'-t'')}\nonumber\\
	\label{eq:Fpp:BpIntEq}
	\sB'_t 	& = \int_t^\infty\! dt \int_{t'}^\infty \! dt''\;
					e^{\sR(t'-t'')}\sK_{t''}e^{-\sL t'}\sD\\
				& \quad  - \int_0^t \! dt' \int_{t-t'}^\infty \! dt'' \; 
					e^{\sR(t-t'-t'')}\sK_{t''}\sB'_{t'}e^{-\sR(t-t')}.\nonumber
\end{align}
\end{subequations}
Applying the matrix norm $\nrm{\cdot}$ to both sides of these equations and using the conditions \eqref{eq:Ftt:kernelbound} and $\nrm{\sL},\nrm{\sR}\leq\rho$ yields the implicit bound 
\begin{align}
	\nrm{\sX'_t} & \leq \frac{M}{(k-\rho)^2}\nrm{\sD}e^{-(k-\rho)t}\\
		&\quad + \frac{M}{k-\rho}\tint\nrm{\sX'_{t'}}e^{-(k-\rho)(t-t')},\nonumber
\end{align}
which, upon defining $\xi_t = \nrm{\sX'_t}e^{(k-\rho)t}$, becomes
\begin{equation}
	\xi_t \leq \frac{M}{(k-\rho)^2}\nrm{\sD} + \frac{M}{k-\rho}\tint\xi_{t'}.
\end{equation}
This bound can be made explicit by means of Gr\"onwall's inequality \cite{pachpatte1997}, which implies 
\begin{equation}
	\xi_t \leq \frac{M}{(k-\rho)^2}\nrm{\sD} e^{Mt/(k-\rho)}. 
\end{equation}
Hence, we have 
\begin{equation}
	\nrm{\sX'_t} \leq \frac{M}{(k-\rho)^2}\nrm{\sD}e^{-(\eta-\rho)t},
\end{equation}
where we have used the definition \eqref{eq:Ftt:defeta} of $\eta$. 
Since $\eta > \rho$, it follows that $\nrm{\sX'_t}\rightarrow 0$ for $t\rightarrow\infty$, that is, 
\begin{equation}
	\lim_{t\rightarrow\infty} \nrm{\sA_t-\sD} = \lim_{t\rightarrow\infty}\nrm{\sB_t-\sD} = 0.
\end{equation}
Consequently, we have shown that the reduced propagators $\sA_t$ and $\sB_t$ both converge to the matrix $\sD$ in the long-time limit. 

\subsubsection{Long-Time Approximation}\label{sec:FpfLTA}

To derive the bound \eqref{eq:Ftt:LTAErr} on the error of the long-time approximation, we note that, within the weak-memory regime, the propagator is given by $\sZ_t=e^{\sL t}\sA_t = \sB_t e^{\sR t}$ and thus satisfies the differential equations  
\begin{equation}\label{eq:Fpa:effleeq}
	\dot{\sZ}_t = \sL \sZ_t + \sE_t \quad\text{and}\quad
	\dot{\sZ}_t = \sZ_t \sR + \sF_t
\end{equation}
with the matrix functions  
\begin{equation}\label{eq:Fpa:EFDef} 
	\sE_t = e^{\sL t}\dot{\sA}_t \quad\text{and}\quad
	\sF_t = \dot{\sB}_t e^{\sR t},
\end{equation}
to which we will henceforth refer as memory functions. 
Our first aim is to derive a bound on these objects. 
To this end, we differentiate the Eqs.~\eqref{eq:Fpp:ApIntEq} and \eqref{eq:Fpp:BpIntEq} with respect to $t$ and note that $\dot{\sX}'_t = \dot{\sX}_t$ for $\sX = \sA,\sB$. 
Upon inserting the definitions \eqref{eq:Fpa:EFDef} of $\sE_t$ and $\sF_t$, respectively, we obtain the integral equations 
\begin{subequations}
\begin{align}
	\sE_t & = - \int_t^\infty \! dt' \; \sK_{t'}e^{\sL(t-t')}\\
		  & \quad  - \int_0^t \! dt' \int_{t-t'}^\infty \! dt'' \; 
		  	\sE_{t'}\sK_{t''}e^{\sL(t-t'-t'')}\nonumber\\
	\sF_t & = - \int_t^\infty \! dt' \; e^{\sR(t-t')}\sK_{t'}\\
		  & \quad - \int_0^t \! dt' \int_{t-t'}^\infty \! dt'' \;
		  	e^{\sR (t-t'-t'')}\sK_{t''}\sF_{t'}.\nonumber
\end{align}
\end{subequations}
Applying the norm $\nrm{\cdot}$ to both sides of these equations and again using the conditions \eqref{eq:Ftt:kernelbound} and $\nrm{\sL},\nrm{\sR}\leq\rho$ yields the implicit bound 
\begin{equation}
	\zeta_t \leq k-\eta + (k-\eta)\tint \zeta_{t'},
\end{equation}
where $\eta$ was defined in Eq.~\eqref{eq:Ftt:defeta} and $\zeta_t = \nrm{\sY_t}e^{kt}$ with $\sY=\sE,\sF$.
Upon using Gr\"onwall's inequality to make this bound explicit \cite{pachpatte1997}, we obtain the bound
\begin{equation}\label{eq:Fpa:EFBnd}
	\nrm{\sE_t},\nrm{\sF_t} \leq (k-\eta)e^{-\eta t}. 
\end{equation}
This result implies that the solution $x_t$ of the non-local time evolution equation \eqref{eq:int:nleeq} satisfies the quasi-local equation \eqref{eq:int:leeq} with an exponentially decaying inhomogeneity, as was stated in the introduction. 

We now observe that the general solutions of the differential equations \eqref{eq:Fpa:effleeq} are 
\begin{subequations}\label{eq:Fpa:LRansatz}
\begin{align}
	\label{eq:Fpa:Lansatz}
	\sZ_t & = e^{\sL t} + \tint e^{\sL(t- t')}\sE_{t'},\\
	\label{eq:Fpa:Ransatz}
	\sZ_t & = e^{\sR t} + \tint \sF_{t'} e^{\sR(t-t')}.
\end{align} 
\end{subequations}
Hence, the slippage matrix $\sD$ must be given by 
\begin{align}
	\label{eq:Fpa:DModE}
	\sD & 	= \lim_{t\rightarrow\infty} e^{-\sL t}\sZ_t 
			= \mathsf{1} + \fint e^{-\sL t} \sE_t\\
	\label{eq:Fpa:DModF}
		& 	= \lim_{t\rightarrow\infty} \sZ_t e^{-\sR t} 
			= \mathsf{1} + \fint \sF_t e^{-\sR t},
			\nonumber
\end{align}
where the bounds \eqref{eq:Fpa:EFBnd} and $\nrm{\sL},\nrm{\sR}\leq \rho$ ensure that the improper integrals are well defined. 
Upon recalling the definition \eqref{eq:Ftt:LTA} of the long-time approximation and inserting $x_t = \sZ_t x_0$, we thus find  
\begin{align}
	\abs{x_t - y_t} &=\abs{\int_t^\infty\! dt'\;e^{\sL(t-t')}\sE_{t'} x_0}\\
	& = \abs{\int_t^\infty\! dt'\;\sF_{t'}e^{\sL(t-t')} x_0} \nonumber\\
	&\leq (k-\eta) \abs{x_0} \int_t^\infty\!dt'\; e^{\rho(t'-t)}e^{-\eta t'}\nonumber\\
	& = \frac{k-\eta}{\eta-\rho}\abs{x_0} e^{-\eta t}, \nonumber
\end{align}
where we have used that the vector norm $\abs{\cdot}$ is consistent with the matrix norm $\nrm{\cdot}$. 
Hence, we have shown that there exists at least one set of generators and reduced propagators, for which Theorem~\ref{thr:LG} holds. 

\subsubsection{Uniqueness}\label{sec:FpfUni}

We take $\sL$ and $\sR$ to be the generators identified in Sec.~\ref{sec:FpfGen} and denote by $\sA_t$ and $\sB_t$ the corresponding reduced propagators. 
We now assume that there exists a second set of generators $\sL'$ and $\sR'$ and reduced propagators $\sA'_t$ and $\sB'_t$ such that 
\begin{equation}
	\sZ_t = e^{\sL' t}\sA'_t = \sB'_t e^{\sR't},
\end{equation}
where $\sA'_t$ and $\sB'_t$ converge to some non-singular matrices $\sA'_\infty$ and $\sB'_\infty$ in the limit $t\rightarrow\infty$. 
Since the propagator is unique \cite{burton2005}, we then have
\begin{equation}\label{eq:Fpu:prop}
	e^{\sL' t} \sA'_t = e^{\sL t}\sA_t,
		\quad
		\sB'_t e^{\sR't} = \sB_t e^{\sR t}.
\end{equation}
Furthermore, since $\sA_t$ and $\sB_t$ become non-singular for $t\rightarrow\infty$, there has to exist a $t_1\geq 0$ such that $\sA_t$ and $\sB_t$ are non-singular for any $t\geq t_1$. 
Hence, for sufficiently large $t$, we can rearrange the identities \eqref{eq:Fpu:prop} to 
\begin{equation}\label{eq:Fpu:STDef}
	e^{-\sL't}e^{\sL t} = \sA'_t \sA_t^{-1} = \sS_t, 
		\quad
		e^{\sR t}e^{-\sR't} = \sB^{-1}_t \sB'_t = \sT_t,
\end{equation}
where $\sS_t$ and $\sT_t$ converge to the non-singular matrices
\begin{equation}\label{eq:Fpu:STAsymp}
	\sS = \sA'_\infty\sD^{-1},
		\quad
		\sT = \sD^{-1}\sB'_\infty.
\end{equation}
At the same time, these matrix functions satisfy the differential equations 
\begin{equation}
	\dot{\sS}_t = \sS_t\sL - \sL'\sS_t,
		\quad
		\dot{\sT}_t = \sR\sT_t - \sT_t \sR', 
\end{equation}
which show that $\dot{\sS}_t$ and $\dot{\sT}_t$ have well defined limits for $t\rightarrow\infty$.  
Since these limits exist, they must be zero, which in turn is only possible if
\begin{equation}\label{eq:Fpu:sim} 
	\sL' = \sS\sL\sS^{-1},
		\quad
		\sR' = \sT^{-1} \sR \sT. 
\end{equation}
This result shows that the generators are unique up to similarity transformations. 

We now insert the identities \eqref{eq:Fpu:sim} back into the definitions \eqref{eq:Fpu:STDef} of $\sS_t$ and $\sT_t$ and take the limit $t\rightarrow\infty$, which yields the asymptotic relations 
\begin{equation}\label{eq:Fpu:STasymp}
	\lim_{t\rightarrow\infty} e^{-\sL t}\sS^{-1}e^{\sL t} 
		=\mathsf{1},
		\quad
		\lim_{t\rightarrow\infty} e^{\sR t} \sT^{-1} e^{-\sR t} 
		=\mathsf{1}. 
\end{equation}
To determine whether there exist non-trivial matrices $\sS^{-1}$, $\sT^{-1}\neq\mathsf{1}$ that satisfy these conditions, we can, without loss of generality, assume that the generators have Jordan normal form \cite{horn2013},
\begin{equation}
	\sL = \bigoplus_i (\lambda_i + \sN_i), 
		\quad
		\sR = \bigoplus_i (\varrho_i + \sN_i).
\end{equation}
Here, $\lambda_i, \varrho_i\in\bC$ are the eigenvalues of $\sL$ and $\sR$, which we assume to be ordered by decreasing real part, and the $\sN_i\in\bC^{N_i\times N_i}$ are nilpotent matrix blocks.
Partitioning both sides of the relations \eqref{eq:Fpu:STasymp} conformally with the Jordan-block structure of $\sL$ and $\sR$ yields 
\begin{subequations}\label{eq:Fpu:AsympBlck}
\begin{align}
	\label{eq:Fpu:AsympBlckS}
	\lim_{t\rightarrow\infty} e^{(\lambda_j-\lambda_i)t} e^{-\sN_i t}(\sS^{-1})_{ij} 
			e^{\sN_j t} & = \mathsf{1}_i \delta_{ij},\\
	\label{eq:Fpu:AsympBlckT}
	\lim_{t\rightarrow\infty} e^{(\varrho_i-\varrho_j)t} e^{\sN_i t}(\sT^{-1})_{ij} 
			e^{-\sN_j t} & = \mathsf{1}_i \delta_{ij},
\end{align}
\end{subequations}
where $\mathsf{1}_i$ is the identity matrix with dimension $N_i$.
For these relations to hold, the blocks $(\sS^{-1})_{ij}\in\bC^{N_i\times N_j}$ and $(\sT^{-1})_{ij}\in\bC^{N_i\times N_j}$ must vanish for $j>i$ and $i>j$, respectively,
and the diagonal blocks must be given by $(\sS^{-1})_{ii}=(\sT^{-1})_{ii} = \mathsf{1}_i$.
Hence, $\sS^{-1}$ and $\sT^{-1}$ must be upper and lower triangular matrices, respectively, and all of their eigenvalues must be $1$.
The remaining blocks $(\sS^{-1})_{ij}$ with $j<i$ must still vanish if $\Re{\lambda_j -\lambda_i}=0$; analogously, the blocks $(\sT^{-1})_{ij}$ with $i<j$ must vanish if $\Re{\varrho_i -\varrho_j}=0$.
If the above conditions on the eigenvalues are not met, however, these blocks are arbitrary. 
That is, if the original generators $\sL$ and $\sR$ have at least two eigenvalues with distinct real parts, there exist continuous families of non-singular matrices $\sS^{-1}$ and $\sT^{-1}$ that satisfy the relations \eqref{eq:Fpu:STasymp}.
From any of these matrices, one can construct modified generators $\sL'$ and $\sR'$ such that the corresponding reduced propagators, 
\begin{equation}
	\sA'_t = e^{-\sL' t}\sZ_t,
		\quad
		\sB'_t = \sZ_t e^{-\sR't},
\end{equation}
converge to non-singular matrices, $\sA'_\infty = \sS\sD$, $\sB'_\infty = \sD\sT$, as can be verified by solving the Eqs.~\eqref{eq:Fpu:prop} for $\sA'_t$ and $\sB'_t$ and inserting the identities \eqref{eq:Fpu:sim} to eliminate $\sL'$ and $\sR'$. 

The above analysis shows that requiring the reduced propagators to converge to non-singular matrices is not sufficient to specify a unique set of generators.  
However, if we further require that $\sA'_\infty = \sB'_\infty$, we obtain the additional constraint $\sS\sD = \sD\sT$, which, upon insertion into Eq.~\eqref{eq:Fpu:STasymp}, yields the asymptotic relations
\begin{equation}\label{eq:Fpu:STasympAdd}
	\lim_{t\rightarrow\infty} e^{\sL t}\sS^{-1}e^{-\sL t} 
		=\mathsf{1},
		\quad
		\lim_{t\rightarrow\infty} e^{-\sR t} \sT^{-1} e^{\sR t} 
		=\mathsf{1},
\end{equation}
where we have used that $\sL\sD = \sD\sR$.
Expanding the generators into Jordan blocks and partitioning $\sS^{-1}$ and $\sT^{-1}$ conformally now leads to the conditions
\begin{subequations}
\begin{align}
	\lim_{t\rightarrow\infty} e^{(\lambda_i-\lambda_j)t} e^{\sN_i t}(\sS^{-1})_{ij} 
			e^{-\sN_j t} & = \mathsf{1}_i \delta_{ij},\\
	\lim_{t\rightarrow\infty} e^{(\varrho_j-\varrho_i)t} e^{-\sN_i t}(\sT^{-1})_{ij} 
			e^{\sN_j t} & = \mathsf{1}_i \delta_{ij},
\end{align}
\end{subequations}
which cannot be satisfied simultaneously with those provided by the Eqs.~\eqref{eq:Fpu:AsympBlck}, unless $(\sS^{-1})_{ij}=(\sT^{-1})_{ij}=\mathsf{1}_i\delta_{ij}$, that is, $\sS=\sT=\mathsf{1}$. 
We can thus conclude that there exists exactly one pair of generators $\sL$ and $\sR$ such that 
\begin{equation}\label{eq:Fpu:PropGen}
	\lim_{t\rightarrow\infty} e^{-\sL t}\sZ_t 
		= \lim_{t\rightarrow\infty} \sZ_t e^{-\sR t} = \sD
\end{equation}
with $\sD$ being a non-singular matrix. 
The proof of Theorem~\ref{thr:LG} is therefore complete. 

\subsubsection{Alternative Criteria for Uniqueness}\label{sec:FpfDis}

As we have just seen, the condition \eqref{eq:Fpu:PropGen} defines a unique pair of local left and right generators, which satisfy the intertwining relation \eqref{eq:Ftt:intertw}.
In the following, we refer to these objects as proper generators whenever we need to distinguish them explicitly. 
By contrast, the weaker conditions 
\begin{equation}\label{eq:Fpd:ModGen}
	\lim_{t\rightarrow\infty} e^{-\sL t}\sZ_t = \sA_\infty
		\quad\text{and}\quad
		\lim_{t\rightarrow\infty} \sZ_t e^{-\sR t} = \sB_\infty
\end{equation}
with $\sA_\infty$ and $\sB_\infty$ being non-singular but not necessarily identical, define whole classes of modified left and right generators, which are connected among themselves, but in general not with each other, through similarity transformations. 
Since it might at first appear somewhat arbitrary to enforce uniqueness of the local generators through the  condition $\sA_\infty = \sB_\infty = \sD$, we here provide three alternative characterizations of the proper generators, which are independent from the one given above.   

\textbf{\emph{(i) Fixed-point equations.}} 
The proper generators are unique in that they satisfy both the fixed-point equations \eqref{eq:Fpg:genAsmp} and the bounds $\nrm{\sL},\nrm{\sR}\leq \rho$. 
This criterion follows directly from the results of Sec.~\ref{sec:FpfGen}.

\textbf{\emph{(ii) Long-time approximation.}}
The second part of Theorem~\ref{thr:LG} introduces the long-time approximation $y_t$ of the solution $x_t$ of Eq.~\eqref{eq:FA:nleeq}. 
The corresponding long-time approximation of the propagator $\sZ_t$ is given by  
\begin{equation}
	\sU_t = e^{\sL t}\sD = \sD e^{\sR t},
\end{equation}
where $\sL$ and $\sR$ are the proper generators. 
Using the results of Sec.~\ref{sec:FpfLTA}, it is straightforward to shown that this approximation satisfies 
\begin{equation}\label{eq:Fpu:LTA}
	\nrm{\sZ_t - \sU_t} \leq \frac{k-\eta}{\eta-\rho}e^{-\eta t}.
\end{equation}
For any pair of modified generators $\sL'\neq\sL$ and $\sR'\neq\sR$, we can define the modified long-time approximations 
\begin{equation}
	\hat{\sU}_t = e^{\sL't}\sA'_\infty \quad\text{and}\quad \check{\sU}_t = \sB'_\infty e^{\sR' t},
\end{equation}
which are in general not identical. 
A key observation is now that neither of these approximations can satisfy the bound \eqref{eq:Fpu:LTA}. 
Instead, the asymptotic relations
\begin{equation}\label{eq:Fpu:LTAAsymp}
	\lim_{t\rightarrow\infty} \nrm{\sZ_t - \hat{\sU}_t}e^{\sigma t} 
		=\lim_{t\rightarrow\infty} \nrm{\sZ_t - \check{\sU}_t}e^{\sigma t} = \infty
\end{equation}
hold for any $\sigma>\rho$.
Hence, $\hat{\sU}_t$ and $\check{\sU}_t$ do either not converge to the actual propagator, or at least converge significantly slower than $\sU_t$. 
That is, the proper generators give rise to the unique optimal long-time approximation of the propagator.

To verify this result, we first note that 
\begin{subequations}
\begin{align}
	\label{eq:Fpu:ZSigRela}
	\sZ_t - \hat{\sU}_t & = \sZ_t -\sU_t + (\mathsf{1}-\sS)e^{\sL t}\sD  ,\\
	\label{eq:Fpu:ZSigRelb}
	\sZ_t - \check{\sU}_t & = \sZ_t -\sU_t + \sD e^{\sR t}(\mathsf{1}-\sT),
\end{align}
\end{subequations}
where $\sL$ and $\sR$ are the proper generators and we have used the relations \eqref{eq:Fpu:STAsymp} and \eqref{eq:Fpu:sim}. 
Next, we recall that the eigenvalues of $\sL$ satisfy $\abs{\lambda_i}\leq \rho$. 
Since $\sD$ is non-singular and $\sS\neq \mathsf{1}$ by assumption, there must exist a constant $C>0$ such that 
\begin{equation}
	\nrm{(\mathsf{1}-\sS)e^{\sL t}\sD}\geq C e^{-\rho t}.
\end{equation}
Together with the bound \eqref{eq:Fpu:LTA}, this result implies the existence of a $t_1\geq 0$ such that 
\begin{equation}
	\nrm{(\mathsf{1}-\sS)e^{\sL t}\sD}\geq\nrm{\sZ_t- \sU_t}
\end{equation}
for any $t\geq t_1$. 
Consequently, the lower bound
\begin{align}
	\nrm{\sZ_t-\hat{\sU}_t} & \geq \bigl|\nrm{\sZ_t -\sU_t} - \nrm{(\mathsf{1}-\sS)e^{\sL t}\sD}\bigr|\\
		& \geq C e^{-\rho t} - \frac{k-\eta}{\eta- \rho} e^{-\eta t}\nonumber
\end{align}
holds for any sufficiently large $t$. 
This result shows that $\lim_{t\rightarrow\infty}\nrm{\sZ_t - \hat{\sU}_t}e^{\sigma t} = \infty$ for any $\sigma>\rho$. 
The second relation in Eq.~\eqref{eq:Fpu:LTAAsymp} can be confirmed analogously.

\textbf{\emph{(iii) Memory functions.}}
Finally, the proper generators can be uniquely defined in terms of the asymptotic behavior of their associated memory functions $\sE_t$ and $\sF_t$, which appear in the quasi-local time-evolution equations \eqref{eq:int:effeq} and \eqref{eq:Fpa:effleeq}. 
In general, it is possible to construct unique memory functions for any given generators so that the propagator satisfies the differential equations \eqref{eq:Fpa:effleeq}.
However, as show in Sec.~\ref{sec:Fpt}, the bounds \eqref{eq:Fpa:EFBnd} only hold for the proper memory functions, i.e., those that correspond to the proper generators. 
For any other choices of generators, the Eqs.~\eqref{eq:Fpa:effleeq} can only be satisfied with memory functions $\sE'_t$ and $\sF'_t$ that fulfill the asymptotic relations
\begin{equation}\label{eq:Fpu:NonPropGen}
	\limsup_{t\rightarrow\infty} \; \nrm{\sE'_t} e^{\sigma t} 
		= \limsup_{t\rightarrow\infty} \; \nrm{\sF'_t} e^{\sigma t} = \infty
\end{equation}
for any $\sigma>\rho$. 
That is, $\sE'_t$ and $\sF'_t$ either do not vanish in the limit $t\rightarrow\infty$, or at least decay significantly slower than the proper memory functions. 
In this sense, the proper generators render the Eqs.~\eqref{eq:Fpa:effleeq}, which are equivalent to the non-local time evolution equations \eqref{eq:Fpg:fbwrdeq}, maximally local in time; for an illustration of this result, which we derive in the next section, see Sec.~\ref{sec:Esm}. 

\vspace*{21pt}
\subsubsection{TCL Method Revisited}\label{sec:FpfTCL} 
We conclude this part of our analysis by briefly revisiting the TCL method mentioned in the introduction. 
The time dependent TCL generator is formally given by 
\begin{equation}\label{eq:Fpu:TCLDef}
	\sL_t^{\vphantom{1}} = \dot{\sZ}_t^{\vphantom{1}}\sZ_t^{-1}.
\end{equation}
Since the propagator $\sZ_t$ is unique, so is $\sL_t$. 
If we now set $\sZ_t = e^{\sL t} \sA_t$, where $\sL$ is the proper left generator, we obtain the result   
\begin{equation}\label{eq:Fpu:TCL}
	\sL_t^{\vphantom{1}} = \sL + e^{\sL t} \dot{\sA}_t^{\vphantom{1}}\sA_t^{-1} e^{-\sL t}
\end{equation}
and it can be easily checked that this expression is indeed invariant if $\sL$ and $\sA_t$ are replaced with $\sL' = \sS \sL \sS^{-1}$ and $\sA'_t = \sS e^{-\sL t}\sS^{-1}e^{\sL t}\sA_t$, respectively; 
that is, the proper and any modified left generator all give rise to the same TCL generator.
Upon inserting the definition \eqref{eq:Fpa:EFDef} of $\sE_t$,  Eq.~\eqref{eq:Fpu:TCL} becomes 
\begin{equation}
	\sL_t^{\vphantom{1}} = \sL + \sE_t^{\vphantom{1}}\sA_t^{-1}e^{-\sL t}.
\end{equation}
Since the reduced propagator $\sA_t$ converges to the non-singular matrix $\sD$ in the limit $t\rightarrow\infty$, the bounds \eqref{eq:Fpa:EFBnd} and $\nrm{\sL}\leq\rho$ show that the time dependent part of $\sL_t$ vanishes for $t\rightarrow\infty$. 
Hence, within the weak-memory regime, the TCL generator $\sL_t$ converges to the proper left generator $\sL$ in the long-time limit. 

In addition, the following result holds for finite times. 
By recalling the expression \eqref{eq:Fpa:Lansatz} for the propagator, and again using the bounds \eqref{eq:Fpa:EFBnd} and $\nrm{\sL}\leq\rho$, we find that 
\begin{align}
	\nrm{\sA_t -\mathsf{1}} & = \nrm{\tint e^{-\sL t'} \sE_{t'}} \\
		& \leq \frac{k-\eta}{\eta-\rho} 
		  = \frac{1-\sqrt{1-\varepsilon}}{2\sqrt{1-\varepsilon}}, \nonumber
\end{align}
where we have inserted the definition \eqref{eq:Ftt:parabounds} of $\varepsilon$. 
Hence, we have $\nrm{\sA_t -\mathsf{1}}< 1 $ for $\varepsilon<8/9$. 
Under this condition, the reduced and thus the full propagator are non-singular for any time $t\geq 0$ and the TCL generator, as defined in Eq.~\eqref{eq:Fpu:TCLDef}, is bounded over any finite time interval. 

\subsection{Perturbation Theory}\label{sec:Fpt}

Theorem~\ref{thr:LG} characterizes the general structure of the solutions of Eq.~\eqref{eq:Ftt:nleeq} in the weak-memory regime. 
However, the generators $\sL$ and $\sR$ and the corresponding reduced propagators $\sA_t$ and $\sB_t$ can usually not be determined exactly for non-trivial models. 
It then becomes necessary to resort to approximation methods, which we develop next. 
We first note that the generators can, at least in principle, be found by solving the non-linear matrix equations \eqref{eq:Fpg:genAsmp} through fixed-point iteration, where the initial guesses must be drawn from the set $B_\rho$ defined in Eq.~\eqref{eq:Fpg:defBrho}. 
Once the generators have been determined to sufficient accuracy, the slippage matrix $\sD$ can be obtained from Eq.~\eqref{eq:Fpg:DDef}.
This method gives access to the long-time approximation \eqref{eq:Ftt:LTA}, but not to the short-time evolution of the system.
In the following, we develop an alternative approach, where the generators are determined indirectly through their associated memory functions $\sE_t$ and $\sF_t$. 
To this end, we derive linear fixed-point equations for these objects, which can be solved by iteration to arbitrary accuracy. 
In this way, we obtain series representations of both memory functions and generators, which 
correspond to expansions in the memory strength $\varphi=M/k^2$. 
Truncating these expansions at increasingly higher orders yields successive approximations of the solutions of Eq.~\eqref{eq:Ftt:nleeq}, which converge uniformly on any finite time interval. 
In addition, we confirm that the proper generators are unique in that they correspond to minimal memory functions. 
Readers who have skipped the proof of Theorem~\ref{thr:LG} may go back to Sec.~\ref{sec:FpfDis} for the definitions of proper generators and memory functions.

\subsubsection{Memory Functions}\label{sec:FptMef}

Inserting the general solutions \eqref{eq:Fpa:Lansatz} and \eqref{eq:Fpa:Ransatz} of the Eqs.~\eqref{eq:Fpa:effleeq} into the non-local time-evolution equations \eqref{eq:Fpg:bwrdeq} and \eqref{eq:Fpg:fwrdeq}, respectively, shows that the memory functions $\sE_t$ and $\sF_t$ have to satisfy the linear integro-differential equations 
\begin{subequations}\label{eq:Fbs:EFEq}
\begin{align}
	\label{eq:Fbs:EEq}
	\dot{\sE}_t & = \sK_t + \sE_t\sV + \tint \sE_{t'}\sK_{t-t'},\\
	\label{eq:Fbs:FEq}
	\dot{\sF}_t & = \sK_t + \sV\sF_t + \tint \sK_{t-t'}\sF_{t'} 
\end{align}
\end{subequations}
with respect to the initial conditions 
\begin{equation}\label{eq:Fbs:GenEF} 
	\sE_0 = \sV - \sL, 
		\quad
		\sF_0 = \sV - \sR.
\end{equation} 
These relations hold without any assumptions on the generators. 
For any given $\sL$ and $\sR$, they constitute initial value problems, which make it possible to construct memory functions $\sE_t$ and $\sF_t$ so that the propagator $\sZ_t$ fulfills the inhomogeneous time evolution equations \eqref{eq:Fpa:effleeq}. 
Since the solutions of the Eqs.~\eqref{eq:Fbs:EFEq} are unique for given $\sE_0$ and $\sF_0$ \cite{burton2005}, they establish a one-to-one correspondence between generators and memory functions. 

If the generators are not given, it is still possible to obtain specific memory functions by imposing asymptotic boundary conditions. 
To this end, we assume that $\sE_t$ and $\sF_t$ are drawn from the set of matrix functions 
\begin{equation}
	\mB_{S,\sigma} 
		= \bigl\{\sX_t :\mathbb{R}^+\rightarrow \bCn \; : \; \nrm{\sX_t} \leq S e^{-\sigma t}\bigr\}.
\end{equation}
Here, $\mathbb{R}^+$ denotes the set of non-negative real numbers and $S$ and $\sigma$ are fixed parameters that satisfy
\begin{equation}\label{eq:Fbs:BParaBnd}
	S\geq k-\sigma \quad\text{and}\quad \rho <\sigma\leq \eta,
\end{equation}
where $\rho$ and $\eta$ were defined in the Eqs.~\eqref{eq:Ftt:defrho} and \eqref{eq:Ftt:defeta}.
Next, we set $\sG_t=\sE_t e^{-\sV t}$ and $\sH_t=e^{-\sV t}\sF_t$ 	and note that these transformed memory functions satisfy the integro-differential equations 
\begin{subequations}\label{eq:Fbs:IntmIDE}
\begin{align}
	\dot{\sG}_t & = \sK_t e^{-\sV t} + \tint \sG_{t'}e^{\sV t'}\sK_{t-t'}e^{-\sV t},\\
	\dot{\sH}_t & = e^{-\sV t} \sK_t + \tint e^{-\sV t}\sK_{t-t'}e^{\sV t'}\sH_{t'}.
\end{align}
\end{subequations}
If $\sE_t,\sF_t\in\mB_{S,\sigma}$, both $\sG_t$ and $\sH_t$ vanish in the limit $t\rightarrow\infty$.
We can then integrate the Eqs.~\eqref{eq:Fbs:IntmIDE} over the time interval $[t,\infty)$ without introducing boundary terms on the left-hand sides. 
Upon returning to the original variables, we thus obtain the identities
\begin{subequations}\label{eq:Fbs:FPEqEF}
\begin{align}
	\label{eq:Fbs:FPEqE}
	\sE_t & = - \int_t^\infty \! dt' \; \Bigl(\sK_{t'} 
	+ \int_0^{t'} \! dt''\; \sE_{t''}\sK_{t'-t''}\Bigr)e^{\sV (t-t')},\\
	\label{eq:Fbs:FPEqF}
	\sF_t & = - \int_t^\infty \! dt' \; e^{\sV (t-t')}\Bigl(\sK_{t'} 
	+ \int_0^{t'} \! dt''\; \sK_{t'-t''}\sF_{t''}\Bigr).
\end{align}
\end{subequations}
These integral equations do not require initial values for $\sE_t$ and $\sF_t$ and are equivalent to the integro-differential equations \eqref{eq:Fbs:EFEq}, as long as we restrict the memory functions to some admissible set $\mB_{S,\sigma}$. 
Once the Eqs.~\eqref{eq:Fbs:FPEqEF} have been solved, the generators $\sL$ and $\sR$ can be obtained from the relations \eqref{eq:Fbs:GenEF}. 
Since the proper memory functions, which we seek to determine, satisfy the bounds \eqref{eq:Fpa:EFBnd}, they are contained in any admissible set $\mB_{S,\sigma}$ and therefore must be solutions of the Eqs.~\eqref{eq:Fbs:FPEqEF}. 
It is however not clear at this point whether these solutions are unique and how they can be obtained systematically. 
These problems can again be addressed with the help of Banach's fixed-point theorem \cite{agarwal2018}.

We define two linear operators $\mT$ and $\mU$ so that the Eqs.~\eqref{eq:Fbs:FPEqE} and \eqref{eq:Fbs:FPEqF} can be written as fixed-point equations, $\sE_t = \mT\sE_t$ and $\sF_t = \mU\sF_t$,  and observe that 
\begin{align}
	\nrm{\mT\sX_t} & = \nrm{\int_t^\infty\! dt' \;
		\Bigl(\sK_{t'} + \int_0^{t'} \! dt'' \; 
			\sX_{t''}\sK_{t'-t''}\Bigr)e^{\sV(t-t')}}\nonumber\\
		\label{eq:Fbs:Closure}
		& \leq \frac{M(k-\sigma-S)e^{-kt} }{(k-\sigma)(k-v)} 
		+  \frac{MS e^{-\sigma t}}{(k-\sigma)(\sigma-v)}
\end{align}
for any $\sX_t\in\mB_{S,\sigma}$, where we have used the bound on the kernel \eqref{eq:Ftt:kernelbound} and the definition $\nrm{\sV}=v$. 
The constraints \eqref{eq:Fbs:BParaBnd} now imply $\nrm{\mT\sX_t}\leq S e^{-\sigma t}$. 
That is, the operator $\mT$ maps $\mB_{S,\sigma}$ into itself and the same result can be confirmed for the operator $\mU$ by repeating the steps above.
Next, we equip the set $\mB_{S,\sigma}$ with the metric 
\begin{equation}\label{eq:Fbs:Metric}
	d_\mu(\sX_t,\sY_t) = \sup_{t\geq 0}\; \nrm{\sX_t - \sY_t}e^{\mu t},
\end{equation}
whereby it becomes a complete metric space if we require that $\mu\leq \sigma$ to ensure that the distance between any two matrix functions $\sX_t,\sY_t\in\mB_{S,\sigma}$ is finite. 
For any such pair, we find, upon taking $\mu>v$, that
\begin{align}
	\label{eq:Fbs:Contr}
	& d_\mu(\mT \sX_t,\mT \sY_t)\\
	& = \sup_{t\geq 0}\; \nrm{\int_t^\infty\! dt'\int_0^{t'}\! dt''\; 
		(\sX_{t''}-\sY_{t''})\sK_{t'-t''}e^{\sV(t-t')}}e^{\mu t}\nonumber\\
	& \leq q_\mu d_\mu(\sX_t,\sY_t),\phantom{\int_0^{t'}}\nonumber
\end{align}
where we have used the relation
\begin{equation}
	\nrm{\sX_t - \sY_t} \leq d_\mu(\sX_t,\sY_t)e^{-\mu t}
\end{equation}
and introduced the contraction factor
\begin{equation}\label{eq:Fbs:ContrFact}
	q_\mu = \frac{M}{(k-\mu)(\mu-v)}.
\end{equation} 
If we now further restrict $\mu$ so that $\rho <\mu<\eta$, we have $q_\mu<1$. 
Hence, the operator $\mT$ is a contraction on $\mB_{S,\sigma}$ and so is $\mU$, as can be verified along the same lines. 
Thus, $\mT$ and $\mU$ each have exactly one fixed point in $\mB_{S,\sigma}$. 

This result shows that, for any parameters $S$ and $\sigma$ that satisfy the constraints \eqref{eq:Fbs:BParaBnd}, the set $\mB_{S,\sigma}$ contains exactly one solution of each of the integral equations \eqref{eq:Fbs:FPEqEF}, or, equivalently, exactly one solution of each of the integro-differential equations \eqref{eq:Fbs:EFEq}. 
These solutions correspond to the proper memory functions $\sE_t$ and $\sF_t$, which can now be determined through fixed-point iteration, along with the proper generators $\sL$ and $\sR$. 
Moreover, for any other generators $\sL'$ and $\sR'$, the solutions $\sE'_t$ and $\sF'_t$ of the initial value problems defined by the Eqs.~\eqref{eq:Fbs:EFEq} and \eqref{eq:Fbs:GenEF} cannot be contained in any admissible set $\mB_{S,\sigma}$. 
That is, for any $\sigma>\rho$ and any $S'>0$, there exists a $t\geq 0$ such that 
\begin{equation}
	\nrm{\sE'_t}, \nrm{\sF'_t} > S'e^{-\sigma t}. 
\end{equation}
Since the solutions of the integro-differential equations \eqref{eq:Fbs:EFEq} are bounded over any finite time interval \cite{burton2005}, this result shows that $\sE'_t$ and $\sF'_t$ have to fulfill the asymptotic relations~\eqref{eq:Fpu:NonPropGen}. 

\subsubsection{Series Expansions}\label{sec:FptSEx}

We are now ready to develop a systematic perturbation theory for the memory functions $\sE_t$ and $\sF_t$ and the generators $\sL$ and $\sR$. 
The following results are direct implications of the analysis provided in Sec.~\ref{sec:FptMef}. 
We first determine $\sE_t$ and $\sF_t$ by fixed-point iteration in $\mB_{S,\sigma}$.
Upon choosing the initial guesses $\sE^0_t = \sF^0_t = 0$, we can set $\sigma=\eta$ and $S=k-\eta$.
Iterating the integral equations \eqref{eq:Fbs:FPEqEF} $n$ times then returns the results
\begin{equation}\label{eq:Fbr:EFExp}
	\sE^n_t = \sum_{m=1}^n \sE^{(m)}_t \quad\text{and}\quad
		\sF^n_t = \sum_{m=1}^n \sF^{(m)}_t,
\end{equation}
where the individual summands are defined through the recursion relations 
\begin{subequations}\label{eq:Fbr:EFRec}
\begin{align}
	\label{eq:Fbr:ERec}
	\sE^{(m)}_t &=-\int_t^\infty \! dt' \int_0^{t'} \! dt''\;\sE^{(m-1)}_{t'-t''}\sK_{t''}e^{\sV(t-t')},\\
	\label{eq:Fbr:FRec}
	\sF^{(m)}_t &= -\int_t^\infty \! dt' \int_0^{t'} \! dt''\;e^{\sV(t-t')}\sK_{t''}\sF^{(m-1)}_{t'-t''}
\end{align}
\end{subequations} 
with the initial conditions
\begin{subequations}\label{eq:Fbr:EFInc}  
\begin{align}
	\label{eq:Fbr:EInc} 
	\sE^{(1)}_t &= -\int_t^\infty \! dt' \; \sK_{t'}e^{\sV (t-t')},\\
	\label{eq:Fbr:FInc} 
	\sF^{(1)}_t &= -\int_t^\infty \! dt' \; e^{\sV (t-t')}\sK_{t'}.
\end{align}
\end{subequations}
For any $n$, the approximations $\sE^n_t$ and $\sF^n_t$ satisfy  
\begin{equation}\label{eq:Fbr:MFApproxBnd}
	\nrm{\sE^n_t},\nrm{\sF^n_t} \leq (k-\eta)e^{-\eta t}.
\end{equation}
In the limit $n\rightarrow\infty$, they converge, in the metric $d_\mu$, to the memory functions $\sE_t$ and $\sF_t$.
As a result, we have  
\begin{align}
	\label{eq:Fbr:MFErrBnd}
	\nrm{\sY_t - \sY^n_t} & \leq d_\mu(\sY_t,\sY^n_t) e^{-\mu t}\\
		& \leq \frac{q_\mu^n}{1-q_\mu} d_\mu(\sY^1_t,\sY^0_t) e^{-\mu t}\nonumber\\
		& \leq \frac{q_\mu^n}{1-q_\mu}\frac{M}{k-v}e^{-\mu t}\nonumber\\
		& \leq \frac{q_\mu^n}{1-q_\mu}\frac{M}{k-v},\nonumber
\end{align}
for $\sY=\sE,\sF$ and $\rho < \mu <\eta$, where the contraction factor $q_\mu$ takes its minimal value $q_{\mu^\ast} = \varepsilon$ for $\mu^\ast=(k+v)/2$.
Since the last bound in Eq.~\eqref{eq:Fbr:MFErrBnd} does not depend on $t$, it follows that the successive approximations $\sE^n_t$ and $\sF^n_t$ converge uniformly over $\mathbb{R}^+$ for $n\rightarrow\infty$.
The generators that correspond to these approximations are given by 
\begin{equation}\label{eq:Fbr:GenApprox} 
	\sL^n = \sV - \sE^n_0 \quad\text{and}\quad
		\sR^n = \sV - \sF^n_0,
\end{equation}
where $\sL^n$ and $\sR^n$ satisfy 
\begin{equation}\label{eq:Fbr:GenApproxBnd} 
	\nrm{\sL^n},\nrm{\sR^n} \leq \rho
\end{equation}
for any $n$, as can be seen directly from the bound \eqref{eq:Fbr:MFApproxBnd}. 
For $n\rightarrow\infty$, the approximations $\sL^n$ and $\sR^n$ converge to the generators $\sL$ and $\sR$,  where 
\begin{equation}\label{eq:Fbr:GenErrBnd}
	\nrm{\sL - \sL^n},\nrm{\sR-\sR^n} \leq \frac{q^n_\mu}{1-q_\mu}\frac{M}{k-v}.
\end{equation}
The adiabatic generator $\sL^0=\sR^0=\sV$ is thereby recovered in zeroth order, while the first-order approximation $\sL^1$ of $\sL$ coincides with the Markov generator \eqref{eq:int:MarkovGen}; for an illustration of these results, see Sec.~\ref{sec:Egl}.

To demonstrate that the series \eqref{eq:Fbr:EFExp} can be regarded as expansions in the memory strength $\varphi=M/k^2$, we introduce the rescaled time $s= kt$ along with the dimensionless variables $\bar{\sV}= \sV/k$, $\bar{\sK}_s= \sK_{s/k}/M$, $\bar{\sE}_s =\sE_{s/k}/k$ and $\bar{\sF}_s=\sF_{s/k}/k$ such that $\nrm{\bar{\sV}}=v/k$ and $\nrm{\bar{\sK}_s}\leq e^{-s}$.
The recursion relations \eqref{eq:Fbr:EFRec} and \eqref{eq:Fbr:EFInc} then become
\begin{subequations}
\begin{align}
	\bar{\sE}^{(m)}_s & = -\varphi \int_s^\infty \! ds'\int_0^{s'} \! ds'' \; 
		\bar{\sE}^{(m-1)}_{s'-s''}\bar{\sK}_{s''}e^{\bar{\sV}(s-s')},\\
	\bar{\sF}^{(m)}_s & = -\varphi \int_s^\infty \! ds'\int_0^{s'} \! ds'' \; 
		e^{\bar{\sV}(s-s')}\bar{\sK}_{s''}\bar{\sF}^{(m-1)}_{s'-s''}
\end{align}
\end{subequations}
and
\begin{subequations}
	\begin{align}
		\bar{\sE}^{(1)}_s & = - \varphi \int_s^\infty \! ds' \; \bar{\sK}_{s'} e^{\bar{\sV}(s-s')},\\
		\bar{\sF}^{(1)}_s & = - \varphi \int_s^\infty \! ds' \; e^{\bar{\sV}(s-s')}\bar{\sK}_{s'}.
	\end{align}
\end{subequations}
Thus, we indeed have $\bar{\sE}^{(m)}_s,\bar{\sF}^{(m)}_s \sim \varphi^m$. 

\subsubsection{Properties and Convergence}\label{sec:FptPrc}

The recursion relations \eqref{eq:Fbr:EFRec} and \eqref{eq:Fbr:EFInc} make it possible to calculate the memory functions $\sE_t$ and $\sF_t$ and the generators $\sL$ and $\sR$ order by order in $\varphi$. 
In every order of this perturbation theory, we obtain two approximations 
\begin{equation}\label{eq:Fbc:ZApprox}
	\hat{\sZ}^n_t = e^{\sL^n t}\sA^n_t 
		\quad\text{and}\quad
		\check{\sZ}^n_t = \sB^n_t e^{\sR^n t} 
\end{equation}
of the propagator, where the approximated reduced propagators are given by 
\begin{subequations}\label{eq:Fbc:ABApprox} 
\begin{align}
	\label{eq:Fbc:AApprox} 
	\sA_t^n & = \mathsf{1} + \tint e^{-\sL^n t'}\sE_{t'}^n,\\
	\label{eq:Fbc:BApprox}
	\sB_t^n & = \mathsf{1} + \tint \sF_{t'}^n e^{-\sR^n t'}.
\end{align}
\end{subequations}
Thus, $\hat{x}_t^n=\hat{\sZ}^n_tx_0$ and $\check{x}^n_t=\check{\sZ}^n_t x_0$ are equally accurate approximations of the solution $x_t$ of the non-local master equation \eqref{eq:Ftt:nleeq}. 
However, due to the non-linear dependence of $\hat{\sZ}_t^n$ and $\check{\sZ}_t^n$ on $\sL^n$ and $\sR^n$, they are in general not identical for finite $n$. 
Still, the long-time approximations 
\begin{equation}\label{eq:Fbc:LTAApprox} 
	\hat{y}^n_t = e^{\sL^n t}\sA^n_\infty x_0
		\quad\text{and}\quad
		\check{y}^n_t = \sB^n_\infty e^{\sR^n t}x_0
\end{equation}
are well-defined for any given $n$, since the approximate memory functions $\sE^n_t$ and $\sF^n_t$ and the approximate generators $\sL^n$ and $\sR^n$ satisfy the same bounds 
\eqref{eq:Fbr:MFApproxBnd} and \eqref{eq:Fbr:GenApproxBnd} as their exact counterparts. 
For the same reason, we have 
\begin{equation}
	\abs{\hat{x}^n_t - \hat{y}^n_t},\abs{\check{x}^n_t - \check{y}^n_t}
			\leq\frac{k-\eta}{\eta-\rho}\abs{x_0}e^{-\eta t},
\end{equation}
as can be verified by repeating the steps at the end of Sec.~\ref{sec:FpfLTA}. 
Hence, the faithfulness of the long-time approximation is preserved in every order of our perturbation theory. 

It remains to show that the approximations $\hat{x}^n_t$ and $\check{x}^n_t$ converge to $x_t$ for $n\rightarrow\infty$. 
To this end, we first consider the reduced propagators and note that 
\begin{align}
	& \nrm{\sA_t -\sA^n_t}\\
	& \leq \fint \nrm{e^{-\sL t}\sE_t - e^{\sL^n t}\sE^n_t}\nonumber\\
	& \leq \fint \nrm{(e^{-\sL t}-e^{-\sL^n t})\sE_t} 
		+ \fint \nrm{e^{-\sL^n t}(\sE_t - \sE^n_t)}\nonumber\\
	& \leq \nrm{\sL-\sL^n}\frac{k-\eta}{(\eta-\rho)^2} + \fint \nrm{\sE_t-\sE^n_t}e^{\rho t}, 
	\nonumber
\end{align}
where we have used the relation \eqref{eq:Fpg:expbnd} as well as the bounds \eqref{eq:Fbr:MFApproxBnd} and \eqref{eq:Fbr:GenApproxBnd}.
Upon inserting the Eqs.~\eqref{eq:Fbr:MFErrBnd} and \eqref{eq:Fbr:GenErrBnd}, we obtain the error bound 
\begin{equation}\label{eq:Fbc:ErrBndA}
	\nrm{\sA_t - \sA^n_t} \leq \frac{q^n_\mu}{1-q_\mu} \alpha_\mu,
\end{equation}
where 
\begin{equation}
	\alpha_\mu = \frac{M(k-\eta)}{(\eta-\rho)^2(k-v)} + \frac{M}{(\mu-\rho)(k-v)}.
\end{equation}
This result, which equally holds for $\nrm{\sB_t-\sB^n_t}$, as can be easily verified by repeating the steps above, shows that the successive approximations of the reduced propagators converge uniformly over $\mathbb{R}^+$. 

We now turn to the full propagator. 
Upon using the result \eqref{eq:Fbc:ErrBndA} and the bound 
\begin{equation}
	\nrm{\sA^n_t} \leq \frac{k-\rho}{\eta-\rho},
\end{equation}
which holds for any $n$ and follows by applying the bounds \eqref{eq:Fbr:MFApproxBnd} and \eqref{eq:Fbr:GenApproxBnd} to the expression \eqref{eq:Fbc:AApprox}, we find that 
\begin{align}
	\nrm{\sZ_t-\hat{\sZ}^n_t} & = \nrm{e^{\sL t} \sA_t - e^{\sL^n t}\sA^n_t}\\
		&\leq \nrm{(e^{\sL t} - e^{\sL^n t})\sA_t} + \nrm{e^{\sL^n t}(\sA_t - \sA^n_t)}\nonumber\\
		&\leq \frac{q^n_\mu}{1-q_\mu}(\beta t + \alpha_\mu)e^{\rho t},\nonumber
\end{align}
where we have introduced the abbreviation
\begin{equation}
	\beta = \frac{M(k-\rho)}{(\eta- \rho)(k-v)}.
\end{equation}
The same result holds for $\nrm{\sZ_t-\check{\sZ}^n_t}$.
Therefore, we have 
\begin{equation}
	\abs{x_t - \hat{x}^n_t},\abs{x_t - \check{x}^n_t}\leq \frac{q^n_\mu}{1-q_\mu}\abs{x_0}	
		(\beta t + \alpha_\mu)e^{\rho t}
\end{equation}
in any vector norm $\abs{\cdot}$ that is consistent with the matrix norm $\nrm{\cdot}$, which shows that $\hat{x}^n_t$ and $\check{x}^n_t$ both converge uniformly to $x_t$ over any finite time interval $[0,t]\subset\mathbb{R}^+$. 
In practice, these bounds are of limited relevance, since they become weak at long times. 
This behavior is generally to be expected, since the generators and their approximations may have positive eigenvalues, which can lead to exponentially growing errors.
In more specific situations, where further conditions can be imposed on the generators, it may however still be possible to obtain stronger error bounds. 

\subsection{Non-Perturbative Methods}\label{sec:Fnp}
The perturbative scheme developed above makes it possible to obtain, in principle arbitrarily accurate, approximations of the full generators and reduced propagators.
If one is only interested in specific eigenvalues or eigenvectors of the generators, however, a more direct approach may be preferable, as we show next.
We denote by $\lb\in\bC$ the eigenvalues of $\sL$ and $\sR$, which must be identical as a result of the intertwining relation \eqref{eq:Ftt:intertw}, and by $L_\lb\in\bC^N$ and $\mR_\lb\in\bC^N$ the corresponding right and left eigenvectors such that
\begin{equation}
	\sL L_\lb = \lb L_\lb
		\quad\text{and}\quad
		\mR_\lb^\ast \sR = \lb \mR_\lb^\ast
\end{equation}
with stars indicating the conjugate transpose. 
Upon acting with these vectors on the fixed-point equations \eqref{eq:Fpg:LAsmp} and \eqref{eq:Fpg:RAsmp}, respectively, we obtain the relations 
\begin{subequations}
\begin{align}
	\label{eq:Fnp:EvEqL}
	[\sV+\hat{\sK}_\lb-\lb]L_\lb & = 0,\\
	\label{eq:Fnp:EvEqR}
		\mR^\ast_\lb[\sV+\hat{\sK}_\lb - \lb] & = 0,
\end{align}
\end{subequations}
where 
\begin{equation}
	\hat{\sK}_\lb = \fint \sK_t e^{-\lb t}
\end{equation}
denotes the Laplace transform of the memory kernel. 
Thus, the eigenvalues of the generators are solutions of the non-linear equation
\begin{equation}\label{eq:Fnp:EV}
	\text{det}[\sV+\hat{\sK}_\lb-\lb] = 0,
\end{equation}
which, since $\nrm{\sL},\nrm{\sR}\leq\rho$, must be contained in the set 
\begin{equation}
	D_\rho = \bigl\{z\in\bC : \abs{z}\leq\rho\bigr\}.
\end{equation}

Notably, the converse of this statement is also true as the following argument shows. 
Assume that the vector $L_\lb\neq 0$ satisfies the relation \eqref{eq:Fnp:EvEqL} for some $\lb\in D_\rho$. 
Acting with this vector on the fixed-point equation \eqref{eq:Fpg:LAsmp} and subtracting the result from Eq.~\eqref{eq:Fnp:EvEqL} yields 
\begin{equation}
	(\lb-\sL)L_\lb  
		=\fint\sK_t(e^{-\lb t}-e^{-\sL t})L_\lb.
\end{equation}
By applying a vector norm $\abs{\cdot}$ that is consistent with the matrix norm $\nrm{\cdot}$ to both sides of this equation and using the formula \eqref{eq:Fpg:expdiff} we find that
\begin{equation}
	\abs{(\lb-\sL)L_\lb} \leq q_\varepsilon \abs{(\lb-\sL)L_\lb},
\end{equation}
where we have inserted the definition \eqref{eq:Fpg:ContFact} of $q_\varepsilon$. 
Since $\varepsilon<1$ and thus $q_\varepsilon <1$, this inequality can only hold if $\abs{(\lb-\sL)L_\lb} =0$. 
Hence, $L_\lb$ must be a right eigenvector of $\sL$ with corresponding eigenvalue $\lb$. 
Along the same lines, it can be shown that any vector $\mR_\lb\neq 0$ that satisfies Eq.~\eqref{eq:Fnp:EvEqR} for some $\lb\in D_\rho$ must be a left eigenvector of $\sR$. 
Thus, any solution $\lb\in D_\rho$ of Eq.~\eqref{eq:Fnp:EV} must correspond to an eigenvalue of the generators. 
Once these eigenvalues have been obtained, the respective right and left eigenvectors of $\sL$ and $\sR$ can be found by solving the linear equations \eqref{eq:Fnp:EvEqL} and \eqref{eq:Fnp:EvEqR}.   

This result provides a direct means of characterizing the long-time behavior of the solution $x_t$ of Eq.~\eqref{eq:Ftt:nleeq}, which is determined by the eigenvalues of the generators. 
For instance, if all solutions of Eq.~\eqref{eq:Fnp:EV} in $D_\rho$ have strictly negative real parts, we have
\begin{equation}
\lim_{t\rightarrow\infty}x_t
	=\lim_{t\rightarrow\infty} e^{\sL t} \sD x_0 
	=\lim_{t\rightarrow\infty} \sD e^{\sR t} x_0 =0
\end{equation} 
for any $x_0$.
Conversely, if there exists at least one solution of Eq.~\eqref{eq:Fnp:EV} in $D_\rho$ with strictly positive real part, then there are initial states for which $x_t$ is unbounded. 
In practice, one is often interested in situations where $x_t$ is bounded for any $x_0$ and, for a certain class of initial states, converges to a non-zero steady state $x_\infty$, which has to satisfy $[\sV+\hat{\sK}_0]x_\infty = 0$, as was already shown in Ref.~\cite{timm2011}. 
To this end, it is necessary that $\lambda=0$ is a solution of Eq.~\eqref{eq:Fnp:EV}, while all remaining solutions in $D_\rho$ must have strictly negative real parts. 
For a sufficient criterion, one further has to verify that $\lambda=0$ is a non-defective eigenvalue, since otherwise $x_t$ may still be unbounded. 
This condition is met if the null eigenvectors of the generators can be chosen such that \cite{horn2013}
\begin{equation}\label{eq:Fnp:Nondef}
	\mL^{j\ast}_0 L^j_0 = \mR^{j\ast}_0 R^j_0 \neq 0
\end{equation}
Here, $L^j_0$ and $\mR^j_0$ are the linearly independent solutions of the Eqs.~\eqref{eq:Fnp:EvEqL} and \eqref{eq:Fnp:EvEqR} for $\lambda=0$ and the associated left and right eigenvectors of $\sL$ and $\sR$, respectively, are given by $\mL^{j\ast}_0 =\mR^{j\ast}_0\sD^{-1}$ and $R_0^j = \sD^{-1}L_0^j$ as a result of the intertwining relation \eqref{eq:Ftt:intertw}. 
Thus, upon inserting the expression \eqref{eq:Fpg:DDef} for the matrix $\sD$, Eq.~\eqref{eq:Fnp:Nondef} becomes 
\begin{equation}
	\mR^{j\ast}_0\sD^{-1}L^j_0 
		= \mR^{j\ast}_0[\mathsf{1}+ \partial_\lb \hat{\sK}_\lb|_{\lb=0} ]L^j_0 \neq 0. 
\end{equation}
If this additional condition is satisfied, $x_t$ is bounded and approaches a non-zero steady state $x_\infty\in\text{span}[\{L^j_0\}]$ whenever $\mR^{j\ast}_0 x_0 \neq 0$ for at least one $\mR^j_0$; otherwise $x_t$ vanishes in the long-time limit. 
If further $\lambda=0$ is a non-degenerate eigenvalue of the generators, that is, if the Eqs.~\eqref{eq:Fnp:EvEqL} and \eqref{eq:Fnp:EvEqR} each have only one linearly independent solution for $\lambda=0$, then the steady state $x_\infty$ is independent of the initial state up to a scaling factor.

\subsection{Extensions}\label{sec:Fex}
So far, we have analyzed the solutions of the non-local time evolution equation \eqref{eq:Ftt:nleeq} under the assumption that that the initial conditions of the system are specified by a given state vector $x_0$ at the time $t=0$. 
We now consider two generalizations of this situation. 
First, we may extend Eq.~\eqref{eq:Ftt:nleeq} by an inhomogeneous term such that 
\begin{equation}\label{eq:Fex:inhnleeq}
	\dot{x}_t = \sV x_t + \tint \sK_{t'}x_{t-t'} + f_t,
\end{equation}
where $f_t\in\bC^N$ is a given vector function. 
If the evolution of the system begins at $t=0$ with a given state $x_0$, the general solution of this equation for $t\geq 0$ is given by the variation of parameters formula \cite{burton2005}
\begin{equation}\label{eq:Fex:VarPara}
	x_t = \sZ_t x_0 + \tint \sZ_{t'} f_{t-t'}.
\end{equation}
Since the propagator $\sZ_t$ follows the inhomogeneous equations of motion \eqref{eq:Fpa:effleeq} in the weak-memory regime, it follows by insertion that $x_t$ can be found by solving the differential equation 
\begin{align}
	\label{eq:Fex:inhlocME}
	\dot{x}_t & = \sL x_t + g_t \quad \text{with}\\
	g_t & = \sE_t x_0 + f_t + \tint \sE_{t'} f_{t-t'}. \nonumber
\end{align}
If the generator and the memory function are not explicitly accessible, approximate solutions of Eq.~\eqref{eq:Fex:inhnleeq} can still be obtained from Eq.~\eqref{eq:Fex:inhlocME} upon replacing $\sL$ and $\sE_t$ with suitable approximations $\sL^n$ and $\sE^n_t$, which can be constructed through the method outlined in Sec.~\ref{sec:Fpt}. 

As a second extension of our framework, we now include situations where the initial conditions of the system are specified in terms of a whole history $\xi_t\in\bC^n$ such that $x_t = \xi_t$ for $t\leq t_0$, where $t_0\geq 0$. 
For $t\geq t_0$, the state vector then follows the time evolution equation 
\begin{equation}
	\dot{x}_t = \sV x_t + \int_{0}^{t_0} \! dt' \; \sK_{t-t'}\xi_{t'} 
		+ \int_{t_0}^{t} \! dt' \; \sK_{t-t'} x_{t'} + f_t. 
\end{equation}
The translated state vector $x'_t = x_{t+t_0}$ therefore satisfies 
\begin{align}
	\label{eq:Fex:inhhistME}
	\dot{x}'_t & = \sV x'_t + \tint \sK_{t'} x'_{t-t'} + f'_t \quad\text{with}\\
	f'_t &= \int_0^{t_0} \! dt' \; \sK_{t+t_0-t'} \xi_{t'} + f_{t+t_0}
	\nonumber
\end{align}
for $t\geq 0$ with the initial condition $x'_0= \xi_{t_0}$ \cite{burton2005}. 
This problem is formally identical to the one discussed in the previous paragraph and therefore can be dealt with in the same way. 
Once Eq.~\eqref{eq:Fex:inhhistME} has been solved, the state vector of the system is given by 
\begin{equation}
	x_t = \begin{cases}
		\xi_t &\text{for}\quad 0\leq t\leq t_0\\
		x'_{t-t_0} &\text{for}\quad t>t_0
	\end{cases}.
\end{equation}
With this result, we conclude the formal part of this paper and move on to applications. 

\section{Applications}\label{sec:E}

\subsection{Coarse Graining of Markov Jump Networks}\label{sec:Emj}

Markov jump processes provide a versatile mathematical framework to model physical systems that admit a discrete state space and are coupled to fast-relaxing environments that do not induce any significant memory effects \cite{vankampen2007}. 
Generically, these conditions apply, for instance, to bio-molecular systems, chemical reaction networks or nano-scale electronic devices in the Coulomb-blockade regime \cite{seifert2012,benenti2017}. 
In many situations, however, only a fraction of the relevant state space is actually accessible.  
The externally distinguishable configurations of complex bio-molecules, for example, may have a complicated internal structure, which is hidden from the observer \cite{hummer2015,martinez2019,hartich2021,ayaz2021,ertel2022,hartich2023,zhao2024}. 
To account for this lack of information, the inaccessible states can be eliminated from the original network. 
Such procedures are known as coarse-graining and inevitably induces a degree of memory into the model, which can be weak if the inaccessible parts of the network evolve faster than the accessible ones. 
In the following,  we show that coarse grained Markov jump networks can be described with non-local time evolution equations of the form \eqref{eq:Ftt:nleeq}. 
The methods developed in the previous section are therefore directly applicable to this class of systems, provided that the weak-memory conditions are satisfied. 

\subsubsection{Markov Jump Processes}\label{sec:EmjMaj}

Formally, a Markov jump network is defined by a discrete set of micro-states $\mathbb{M}=\{i: 1\leq i \leq M \}$ and a set of rates $w(i,j)\geq 0$, which determine the frequency of stochastic transitions from $j$ to $i$. 
On the ensemble level, the state of the network at a given time $t\geq 0$  is described by the probability vector $X_t = [P^1_t,\dots,P^M_t]^\trans$, which assigns an occupation probability $P^i_t$ to every micro-state $i$.
Hence, the elements of $X_t$ are non-negative and sum to $1$. 
The time evolution of the probability vector is governed by the master equation
\begin{equation}\label{eq:Emj:ME}
	\dot{X}_t = \sW X_t.
\end{equation}
The off-diagonal elements of the matrix $\sW$ are given by $(\sW)_{ij} = w(i,j)$ and the diagonal elements are determined by the condition $(1_M)^\trans\sW = 0$, which ensures that $(1_M)^\trans X_t = 1$ for any $t\geq 0$. 
Throughout this section, $1_d$ denotes the vector of dimension $d$ whose entries are all $1$ and the superscript $\trans$ indicates the transpose.

\subsubsection{Coarse Graining}\label{sec:EmjCgr}

A coarse graining of a Markov jump network is induced by a partitioning of the space of micro-states $\bM$ into $N$ disjoint subsets $\bM_\alpha =\{i_\alpha : 1\leq i_\alpha \leq M_\alpha\}$, which are called meso-states \cite{strasberg2019,esposito2012}. 
The aim is now to describe the effective network meso-states without reference to individual micro-states. 
Such a reduction can generally be achieved through standard techniques, which we briefly recapitulate here, following essentially the approach of Ref.~\cite{hummer2015}.
We assume, without loss of generality, that the micro-states are arranged such that 
$X_t = \bigoplus_{\alpha=1}^N X^\alpha_t$, where the vector $X^\alpha_t = [P^{1_\alpha}_t, \dots, P^{M_\alpha}_t]^\trans$ contains the occupation probabilities of all micro-states that belong to the meso-state $\alpha$. 
Next, we divide the rate matrix into a disconnected part $\sW_0 = \bigoplus_\alpha \sW^\alpha$, which describes the internal dynamics of the isolated meso-states, and a connected part $\sW_1 = \sW -\sW_0$, which accounts for transitions between different meso-states. 
Hence, the off-diagonal elements of the partial rate matrix $\sW^\alpha$ are given by $(\sW^\alpha)_{ij} = w(i_\alpha,j_\alpha)$ and the diagonal elements are determined by the condition $(1_{M_\alpha})^\trans\sW^\alpha =0$. 
For simplicity, we assume that each of these matrices is irreducible. 
There then exists a set of unique stationary probability vectors $X^\alpha_\infty$, which satisfy $\sW^\alpha X^\alpha_\infty = 0$. 
From these vectors, we construct a projection matrix  
\begin{equation}\label{eq:Emj:Proj}
	\sP = \bigoplus_\alpha X^\alpha_\infty (1_{M_\alpha})^\trans,
\end{equation}
which makes it possible to formally separate the accessible part of the probability vector, 
$X^0_t = \sP X_t$, from the inaccessible part $X^1_t = \sQ X_t$, where $\sQ = \mathsf{1}-\sP$. 
Upon using the properties $\sP^2 = \sP$ and $\sP\sW_0 = \sW_0\sP =0$ of the projection matrix \eqref{eq:Emj:Proj}, the master equation \eqref{eq:Emj:ME} can be rewritten as a set of two coupled equations of motion,   
\begin{equation}\label{eq:Emj:NZEq}
	\frac{d}{dt}
	\left[\begin{array}{l}
		X^0_t\\ 
		X^1_t
	\end{array}\right]
		= \left[\begin{array}{ll}
			\sP\sW_1 & \sP\sW_1\\
			\sQ\sW_1 & \sQ\sW
		\end{array}\right]
		\left[\begin{array}{l}
				X^0_t\\
				X^1_t
		\end{array}\right]. 
\end{equation}
Eliminating $X^1_t$ from these equations yields a closed time evolution equation for $X^0_t$, which is given by
\begin{equation}
	\dot{X}^0_t = \sP\sW_1 X^0_t 
		+ \tint \sP \sW_1 e^{\sQ\sW t'}
			\sQ\sW_1 X^0_{t-t'}.
\end{equation}
Here, we have set $X^1_0 = 0$. 
That is, we assume the initial micro-state in each meso-state to be drawn from the corresponding stationary distribution $X^\alpha_\infty$, while the total initial occupation probabilities of the meso-states are arbitrary. 
Finally, we define the reduced probability vector $x_t = [p^1_t,\dots,p^N_t]^\trans$, where $p^\alpha_t = (1_{M_\alpha})^\trans X^\alpha_t$ denotes the  occupation probability of the meso-state $\alpha$. 
This vector is linked to $X^0_t$ by the contraction and dilation matrices 
\begin{equation}
	\sM= \bigoplus_\alpha (1_{M_\alpha})^\trans,
		\quad
		\sN = \bigoplus_\alpha X^\alpha_\infty,
\end{equation}
via the relations $\sM X_t^0 = x_t$ and $\sN x_t = X_t^0$.
The time evolution equation \eqref{eq:Emj:NZEq} can thus be cast into the form of the non-local equation \eqref{eq:Ftt:nleeq} with the adiabatic generator
\begin{equation}\label{eq:Emj:RedV}
	\sV = \sM\sP\sW_1 \sN 
\end{equation}
and the memory kernel 
\begin{equation}\label{eq:Emj:RedKer}
	\sK_t = \sM \sP \sW_1 e^{\sQ\sW t}
		\sQ\sW_1\sN. 
\end{equation}
This result is formally exact and holds regardless of how the space of micro-states is initially partitioned into meso-states. 
In practice, however, the coarse graining only reduces the complexity of the initial problem if the memory kernel decays sufficiently fast such that the resulting non-local time evolution equation can be further simplified. 
For this condition to be met, the internal transitions in the meso-states usually have to occur at larger rates than those between different meso-states. 

\subsubsection{Example}\label{sec:EmjExm}

\begin{figure}
\includegraphics[width=8.5cm]{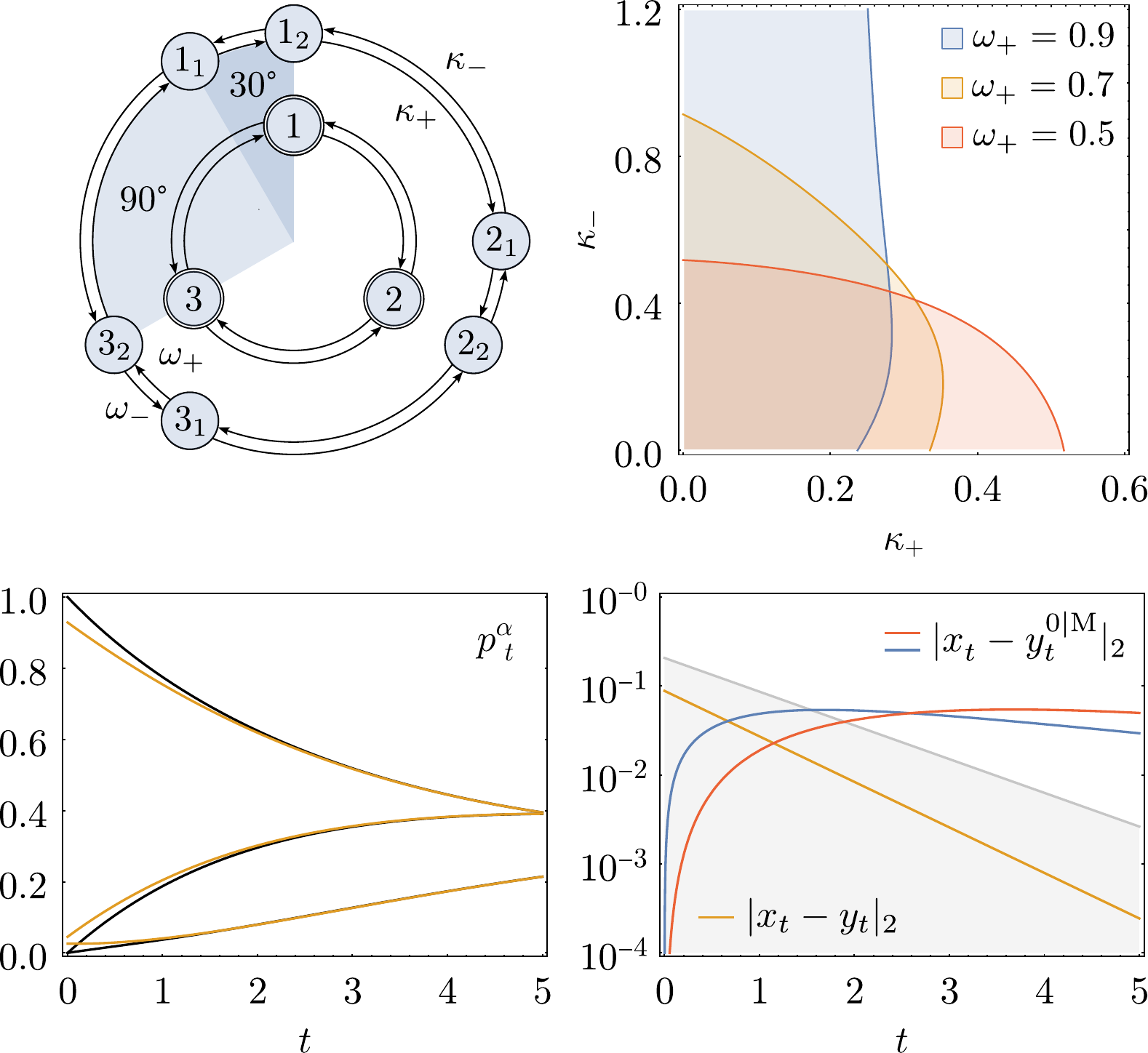}
\caption{F\textsubscript{1}-ATPase as a coarse grained Markov jump network. 
For the plots in the bottom row, we have set $p^\alpha_0 = \delta_{1\alpha}$ and chosen the parameter values $\omega_+=0.8$, $\omega_-=0.2$, $\kappa_+=0.3$ and $\kappa_-=0.5$ such that $v/k\simeq 0.42$ and $\varepsilon=4M/(k-v)^2\simeq 0.89$. 
Since we set $\omega_++\omega_-=1$, all quantities are dimensionless. 
\textbf{Top left:} Schematic representation of the model described in the main text.  
The outer and inner rings shows the original network of micro-states and the effective network obtained by lumping the micro-states $1_\alpha$ and $2_\alpha$ into the meso-state $\alpha$. 
\textbf{Top right:} Weak-memory regime. 
The shaded areas indicate the regions of the $\kappa_+$-$\kappa_-$ space, where the weak memory conditions \eqref{eq:Ftt:parabounds} are satisfied. 
Different colors correspond to different values of $\omega_+$. 
\textbf{Bottom left:} Occupation probabilities of the meso-states $1$, $2$ and $3$, from top to bottom. 
Black lines correspond to the exact solution of the original master equation and orange lines show the long-time approximation \eqref{eq:Emj:LTA}. 
\textbf{Bottom right:} Error of the long-time approximation in the Euclidean norm. 
The orange line shows $\abs{x_t-y_t}_2$ and the shaded area indicates the bound \eqref{eq:Ftt:LTAErr}. 
For comparison, the errors of the adiabatic and Markov approximations, $\abs{x_t - y^0_t}_2$ and $\abs{x_t - y^\text{M}_t}_2$, are shown in blue and red. 
\label{Fig:Emj}}
\end{figure}

The F\textsubscript{1}-ATPase, a molecular motor, which converts chemical energy into rotational motion, provides an instructive example for a Markov jump networks that admits a natural coarse graining \cite{yasuda2001,hartich2023}. 
A full revolution of the motor requires three discrete rotations of 120°, whose end points correspond to the meso-states $\alpha=1,2,3$ of the system. 
Each of these rotations occurs in two consecutive steps of 90° and 30°, which involve the binding of an energy-rich ATP molecule and the release of a lower energetic ADP molecule, respectively. 
Every meso-state $\alpha$ thus contains two micro-states $1_\alpha$ and $2_\alpha$ corresponding to the end points of the 90° and 30° steps, see Fig.~\ref{Fig:Emj}.
We denote the transitions rates between the micro-states by 
\begin{align}
	& w(2_\alpha, 1_\alpha) = \omega_+, 
		&& w(1_\alpha, 2_\alpha)  = \omega_-,\\
			& w(1_{\alpha+1},2_\alpha)  = \kappa_+, 
		 		&& w(2_\alpha,1_{\alpha+1}) = \kappa_-,
		 		\nonumber
\end{align}
where $\omega_+\geq \omega_->0$, $\kappa_+\geq\kappa_->0$ and $i_{3+1} = i_1$. 
Hence, the disconnected and the connected parts of the rate matrix $\sW$ are given by 
\begin{equation}
	\sW_0  = \mathsf{1}_3\otimes
		\left[\begin{array}{rr}
			-\omega_+ & \omega_- \\ \omega_+ & -\omega_-
		\end{array}\right]
\end{equation}
and 
\begin{align}
	\sW_1 = 
	\sS \otimes 
		\left[\begin{array}{ll}
			0 & \kp_+ \\ 0 & 0 
		\end{array}\right]
	+ \sS^\trans \otimes 
		\left[\begin{array}{ll}
			0 & 0 \\ \kp_- & 0 
		\end{array}\right]
	- \mathsf{1}_3 \otimes 
		\left[\begin{array}{ll}
			\kp_-  & 0 \\ 0 & \kp_+
		\end{array}\right],
	\nonumber\\
\end{align}
where we have introduced the matrix 
\begin{equation}
	\sS =	\left[\begin{array}{lll}
				0 & 0 & 1 \\ 1 & 0 & 0 \\ 0 & 1 & 0 
			\end{array}\right]
\end{equation}
and $\mathsf{1}_3$ denotes the identity matrix of dimension $3$. 

We now assume that the 30° rotations occur faster than the 90° steps. 
For convenience, we set $\omega_++\omega_- = 1$, whereby all relevant quantities become dimensionless.
The stationary distributions of the meso-states are then given by $X^\alpha_\infty = [\omega_-, \omega_+]^\trans$.
To eliminate the fast transitions, we first construct the projection matrix \eqref{eq:Emj:Proj} and then evaluate the formulas \eqref{eq:Emj:RedV} and \eqref{eq:Emj:RedKer}, which yields the explicit expressions 
\begin{align}
	\sV   & = \sg_+\sS +\sg_-\sS^\trans-\sg_+-\sg_-,\\[6pt]
	\sK_t & = \bigl(\zeta_+\sS + \zeta_-\sS^\trans -\zeta_+ -\zeta_-\bigr)e^{-(\sV+1+\kappa)t}
\end{align}
for the adiabatic generator and the memory kernel of the effective meso-state network. 
Here,  $\sg_\pm = \kp_\pm\oa_\pm$ and $\kappa = \kp_++\kp_-$ are characteristic system parameters and the coefficients $\zeta_\pm$ are defined as 
\begin{equation}
	\zeta_\pm = 2\sg_\pm^2-\sg_\mp^2 + 2\sg_+\sg_--\sg_\pm \kappa. 
\end{equation}
The spectral norms of $\sV$ and $\sK_t$ are given by 
\begin{align}
	\nrm{\sV}_2 &  = v = \sqrt{3(\sigma_+^2+\sigma_-^2 + \sigma_+\sigma_-)},\\[6pt]
	\nrm{\sK_t}_2  & = M e^{-k t},
\end{align}
where $k=1+\kappa-3(\sigma_+ + \sigma_-)/2$ is the overall decay rate of the memory kernel and $M =  v\sqrt{v^2-\kappa^2 - 2\kappa(1-k)}$ its magnitude. 
Since $\sigma_\pm$ and $\kappa$ vanish in the limit $\kp_\pm\rightarrow 0$, so do $v$ and $M$, while $k$ goes to $1$. 
Thus, by continuity, the weak-memory conditions \eqref{eq:Ftt:parabounds} must be satisfied if the rates $\kappa_\pm$, which describe transitions between different meso-states, are sufficiently small. 
A quantitative analysis shows that these rates can in fact be of the same order as the internal transition rates $\omega_\pm$, see Fig.~\ref{Fig:Emj}.

The local generators $\sL$ and $\sR$ can now be determined by inserting the ansatz 
\begin{equation}\label{eq:Emj:GenAns}
	\sL = \sR = \lb_+\sS + \lb_-\sS^\trans -\lb_+-\lb_-
\end{equation}
into either of the fixed point equations \eqref{eq:Fpg:genAsmp}. 
The resulting matrix equations admits four different solution for the coefficients $\lb_\pm$. 
However, the condition $\nrm{\sL}_2 = \nrm{\sR}_2 \leq \rho$ is satisfied only for 
\begin{equation}
	\lb_\pm = \frac{1+\kappa-\sqrt{12\sg_\mp+2u - (1+\kappa)^2}}{6},
\end{equation}
where $u = \sqrt{16 v^2 +(1+\kappa)^4 + 8(1+\kappa)^2(k-1-\kappa)}$. 
With this result, the slippage matrix $\sD$ can be obtained from Eq.~\eqref{eq:Fpg:DDef}.
Upon using that $\sL$ and $\sR$ are identical and commute with $\sK_t$, we find that
\begin{equation}
	\sD= \bigl[1 + \sK_0[\sL + \sV + 1 +\kappa]^{-2}\bigr]^{-1}.
\end{equation}
We can thus compute the long-time approximation 
\begin{equation}\label{eq:Emj:LTA}
	y_t = e^{\sL t} \sD x_0
\end{equation}
of the reduced probability vector, which we plot together with the exact solution $x_t = \sM e^{\sW t} \sN x_0$ of the original master equation in Fig.~\ref{Fig:Emj}. 
For comparison, we also show the adiabatic and Markov approximations, 
\begin{equation}
	y^0_t = e^{\sV t} x_0 
		\quad\text{and}\quad y^\text{M}_t = e^{\sL^1 t} x_0,
\end{equation}
where the Markov generator is given by
\begin{equation}
	\sL^1 = \sV + [2\sV + 1 + \kappa]^{-1}.
\end{equation}
Notably, both of these approximations incur significantly larger errors than the full long-time approximation \eqref{eq:Emj:LTA}, even at relatively short times. 
Before moving on,  we note that, for the parameter values of Fig.~\ref{Fig:Emj}, the coefficient $\lambda_-\simeq - 0.01$ in the decomposition \eqref{eq:Emj:GenAns} is negative, which shows that the local generators of coarse grained Markov jump networks are not necessarily proper rate matrices. 
As a result, the local time evolution equation $\dot{y}_t = \sL y_t$, which defines the long-time approximation $y_t$ up to the initial condition, can in general not be interpreted as the master equation of an effective Markov jump process. 

\subsection{Semi-Markov Jump Networks}\label{sec:Esm}

Semi-Markov jump processes, which are sometimes also called continuous time random walks \cite{bouchaud1990,masuda2017} can, under certain conditions, arise from the coarse-graining of Markov jump processes \cite{wang2007,martinez2019,teza2020}. 
In addition, they provide a straightforward way to phenomenologically account for memory effects in discrete physical systems when the inaccessible degrees of freedom cannot be modeled explicitly. 
This situation can occur, for example, in solid-state settings with finite-size reservoirs which, in contrast to ideal reservoirs, can induce significant back action when exchanging thermodynamic quantities with the system of interest \cite{schaller2014,brange2018,moreira2023}.
On the ensemble level, semi-Markov jump processes are described by non-local time evolution equations of the form \eqref{eq:Ftt:nleeq}. 
The techniques developed in Sec.~\ref{sec:F} are therefore directly applicable to such models. 
In the following, we apply this approach to a simple toy model from mesoscopic physics, which we use to illustrate the distinction between proper and modified generators discussed in Sec.~\ref{sec:FpfDis}. 

\subsubsection{Semi-Markov Jump Processes}\label{sec:EsmSmj}

A general semi-Markov jump process describes stochastic dynamics on a discrete space of micro-sates $\bN= \{i : 1\leq i \leq N\}$. 
In contrast to a Markov jump processes, this model offers the freedom to specify a whole waiting time distribution $\psi_t(i,j)$ for every transition, rather than only an average transition rate \cite{cinlar1975}.
The individual jumps are thereby still uncorrelated. 
Hence, if the system arrives in the state $j$ at the time $t\geq 0$, then
\begin{equation}
	\Psi_\tau(i,j) = \int_0^\tau \! dt \; \psi_t (i,j)
\end{equation}
is the probability for it to transition to the state $i$ during the time interval $[t,t+\tau]$, regardless of which states were visited earlier. 
Consequently, 
\begin{equation}
	\Phi_\tau(j) = 1 - \sum_{i\neq j} \Psi_\tau(i,j)
\end{equation}
is the probability to remain in the state $j$ until the time $t+\tau$. 
If we assume that every state is eventually left, the cumulative waiting time distributions have to obey the normalization condition 
\begin{equation}
	\sum_{i\neq j} \Psi_\infty(i,j) = 1. 
\end{equation}
The ensemble state of the network is described by a probability vector $x_t = (p^1_t,\dots,p^N_t)$, where $p^i_t$ is the occupation probability of the state $i$ at the time $t$. 
This vector follows the non-local time evolution equation \eqref{eq:Ftt:nleeq}, where
the off-diagonal elements of $\sV$ and $\sK_t$ are implicitly defined through the relation 
\begin{equation}\label{eq:Esm:AdGKerDef}
	\psi_t(i,j) = (\sV)_{ij}\Phi_t(j) + \tint (\sK_{t'})_{ij}\Phi_{t-t'}(j),
\end{equation}
which must usually be solved in Laplace space \cite{feller1964,breuer2009}.
The diagonal elements of $\sV$ and $\sK_t$ are fixed by the conditions $(1_N)^\trans\sV = 0$ and $(1_N)^\trans\sK_t =0$. 
Upon requiring that $\sK_t$ is continuous at $t=0$, both $\sV$ and $\sK_t$ are thus uniquely determined for any given set of waiting time distributions. 
For the special choice 
\begin{equation}
	\psi_t(i,j) = w(i,j)e^{-W(j) t},
			\quad
		 	W(j)=\sum_{i\neq j} w(i,j),
\end{equation}
we recover the Markovian case, that is, Eq.~\eqref{eq:Esm:AdGKerDef} yields $(\sV)_{ij} = w(i,j)$ and $(\sK_t)_{ij} =0$ for $i\neq j$. 

\subsubsection{Example}\label{sec:EsmExm}

We consider a quantum dot, which can accommodate at most one charge carrier due to Coulomb repulsion, in contact with a mesoscopic thermo-chemical reservoir.  
This system admits two micro-states, $0$ and $1$, which correspond to the dot being either empty or occupied, and is described by the probability vector $x_t= [p^1_t,p^0_t]^\trans$. 
Transitions between these states occur through the exchange of charge carriers with the reservoir, whose chemical potential and temperature fluctuate in response to the absorption or emission of carriers.
We model this back action as a semi-Markov process. 
For simplicity, we assume that the waiting times for both absorption and emission events are described by the same distribution 
\begin{equation}\label{eq:Esm:ExWTD}
	\psi_t = \frac{2\gamma\kappa}{\sqrt{\kappa(\kappa-2)}}
		\sinh\bigl[\sqrt{\kappa(\kappa-2)}\gamma t\bigr]e^{-\kappa\gamma t}.
\end{equation}
Here, the rate $\gamma$ sets the overall time scale of the system and the parameter $\kappa\geq 2$, which may be regarded as measure for the size of the reservoir, makes it possible to recover a Markov jump process in the limit $\kappa\rightarrow\infty$, where $\psi_t\rightarrow \gamma e^{-\gamma t}$. 

From the relation \eqref{eq:Esm:AdGKerDef}, we find that the adiabatic generator for this model vanishes, $\sV=0$, while the memory kernel takes the simple form 
\begin{equation}
	\sK_t = 4\kappa\gamma^2  e^{-2\kappa\gamma t} \sH_1 
		 \quad\text{with}\quad
		 \sH_1 = \frac{1}{2}\left[
			\begin{array}{rr}
			-1 & 1 \\ 1 & -1
			\end{array} \right]
\end{equation}
The maximum absolute column sum norm of this kernel is given by $\nrm{\sK_t}_1 = Me^{-kt}$ with $M=4\kappa\gamma^2$ and $k=2\kappa\gamma$. 
Hence, the weak-memory conditions \eqref{eq:Ftt:parabounds} are satisfied for $\kappa>4$. 
The proper left generator can now be obtained from the fixed-point equation \eqref{eq:Fpg:LAsmp} and is given by 
\begin{equation}
	\sL = \rho \sH_1
\end{equation}
with $\rho = \gamma\bigl(\kappa-\sqrt{\kappa(\kappa-4)}\bigr)$. 
Diagonalizing this generator shows that the non-singular matrix 
\begin{equation}
	\sS_\alpha = 1 + \alpha\sH_2
		\quad\text{with}\quad
		\sH_2 = \frac{1}{2}\left[
			\begin{array}{rr}
			-1 & 1 \\ -1 & 1 
			\end{array}\right]
\end{equation}
satisfies the relation \eqref{eq:Fpu:STasymp} for any $\alpha\in\bC$. 
Thus, there exists a continuous family 
\begin{equation}\label{eq:Esm:ModGen}
	\sL_\alpha = \sS_\alpha^{\vphantom{-1}}\sL \sS^{-1}_\alpha = \rho \sH_1 - \alpha\rho\sH_2 
\end{equation}
of modified generators, for which the reduced propagator $\sA_{\alpha,t} = e^{-\sL_\alpha t}\sZ_t$ converges to a non-singular matrix $\sA_{\alpha,\infty}$ in the limit $t\rightarrow\infty$. 
The corresponding modified memory functions $\sE_{\alpha,t}$ can be found by solving the integro-differential equation \eqref{eq:Fbs:EEq} in Laplace space with respect to the initial condition $\sE_{\alpha,0} = - \sL_\alpha$, which yields 
\begin{equation}\label{eq:Esm:ModMemF}
	\sE_{\alpha,t} = -\rho e^{-\eta t}\sH_1 + \frac{\alpha\rho}{\eta-\rho}
	\bigl(\eta e^{-\rho t}-\rho e^{-\eta t} \bigr)\sH_2
\end{equation}
with $\eta = \gamma\bigl(\kappa+\sqrt{\kappa(\kappa-4)}\bigr)$. 
Since $\eta>\rho$, this expression shows that the bound \eqref{eq:Fpa:EFBnd} can indeed only hold for any $t\geq 0$ if $\alpha=0$. 
That is, this bound holds only for the proper memory function $\sE_t=\sE_{0,t}$, which corresponds to the proper left generator $\sL=\sL_0$. 
Furthermore, we immediately see that the modified memory function \eqref{eq:Esm:ModMemF} satisfies the asymptotic relation \eqref{eq:Fpu:NonPropGen} for any $\alpha\neq 0$. 

\begin{figure}
\includegraphics[width=8.5cm]{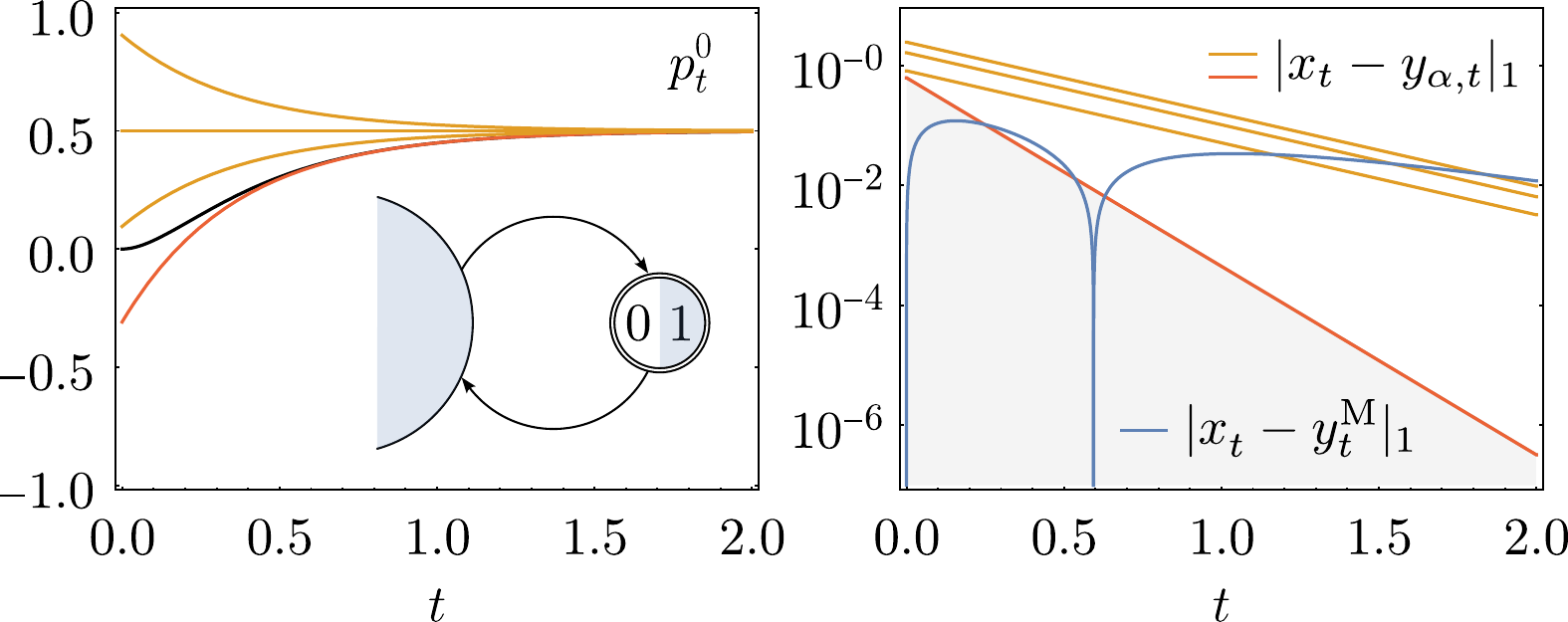}
\caption{\label{Fig:Esm}
Quantum dot in contact with a mesoscopic reservoir. 
For all plots, we have set $x_0=[1,0]^\trans$, $\kappa=5$ and $\gamma=1$. 
Thus, all quantities are dimensionless. 
\textbf{Left:} 
Probability $p^0_t$ for the dot to be empty.  
The inset shows a sketch of the model described in the main text. 
The black line corresponds to the exact solution of Eq.~\eqref{eq:Ftt:nleeq} and the red line shows the proper long-time approximation $y_t=y_{0,t}$. 
Orange lines show the modified long-time approximation \eqref{eq:Esm:ModLTA} for $\alpha=0.5,1.0,1.5$ from bottom to top. 
\textbf{Right:} 
Error of the long-time approximation in the taxicap norm. 
The shaded area indicates the bound \eqref{eq:Ftt:LTAErr}, which is  saturated by the proper long-time approximation shown in red. 
The orange lines show the errors of the modified long-time approximations for $\alpha=0.5,1.0,1.5$ from bottom to top. 
For comparison, we plot the error of the Markov approximation \eqref{eq:Esm:MarGen} in blue. 
} 
\end{figure}

To explore how the transformation \eqref{eq:Esm:ModGen} affects the long-time approximation of the actual dynamics, we note that the matrix $\sA_{\alpha,\infty}$ is given by 
\begin{equation}
	\sA_{\alpha,\infty} = 1 - \frac{\rho}{\eta- \rho}\sH_1 	+ \frac{\alpha\eta}{\eta-\rho}\sH_2
\end{equation}
With this result, which can be obtained from the formula \eqref{eq:Fpa:DModE}, we define the modified long-time approximation 
\begin{equation}\label{eq:Esm:ModLTA}
	y_{\alpha,t} = e^{\sL_\alpha t}\sA_{\alpha,\infty} x_0,
\end{equation}
which is plotted in Fig.~\ref{Fig:Esm}. 
We find that the error of the proper long-time approximation $y_t = y_{0,t}$ exactly saturates the bound \eqref{eq:Ftt:LTAErr}. 
By contrast, the modified long-time approximations incur a significantly larger error;
at intermediate times this error is comparable to the error of the Markov approximation, which is given by  
\begin{equation}\label{eq:Esm:MarGen}
	y^\mathrm{M}_t = e^{\sL^1 t} x_0
	\quad\text{with}\quad
		\sL^1 = 2\gamma \sH_1.
\end{equation}
This behavior is generally expected. 
As we have seen in Sec.~\ref{sec:FpfDis}, on the level of the propagator $\sZ_t$, the error of the proper long-time approximation decays significantly faster than the error of any modified long-time approximation. 
However, on the level of the state vector $x_t$, a strict lower bound on the error of modified long-time approximations, similar to the asymptotic relations \eqref{eq:Fpu:LTAAsymp}, can in general not hold, due to their dependence on the initial state. 
For models with a sufficiently large state space, it is in fact possible to construct modified generators and initial states such that the resulting long-time approximation of the state vector incurs the same error as the corresponding proper long-time approximation.

\subsection{Generalized Langevin Equations}\label{sec:Egl}

The coarse graining procedure described in Sec.~\ref{sec:EmjCgr} provides a special instance of the projection operator method, which has a much wider scope.
In its most general form, this technique can be applied to essentially any dynamical model that admits a separation into relevant and irrelevant parts \cite{fick1990,grabert1992,zwanzig2001,vrugt2020}. 
If the quantity of interest is a vector of relevant observables, the resulting effective equations of motion are referred to as generalized Langevin equations. 
These equations are formally exact and non-local in time. 
Depending on the original model and on how the projection operator is constructed, they can be either linear or non-linear in the observables of interest \cite{zwanzig2001}.
In the former case, they take the form of Eq.~\eqref{eq:Fex:inhnleeq}, where the inhomogeneity corresponds to a generalized stochastic force. 
In practice, the formal expressions for both the stochastic force and the memory kernel can usually not be evaluated explicitly without system specific approximations. 
Nonetheless, the projector operator approach makes it possible to derive a universal relations between these quantities, which is known as the generalized fluctuation dissipation theorem and provides a strong constraint for phenomenological modeling \cite{zwanzig2001}. 

\subsubsection{Setting}\label{sec:EglSet}

We consider a system with relevant observables $O^i_t$ forming the state vector $x_t = [O^1_t,\dots O^N_t]^\trans$, which we assume to satisfy a generalized Langevin equation of the form \eqref{eq:Fex:inhnleeq}. 
The adiabatic generator can then be chosen so that the fluctuating force $f_t$ has vanishing mean. 
If the eliminated degrees of freedom were initially in thermal equilibrium, this force is related to the memory kernel $\sK_t$ by the generalized fluctuation-dissipation theorem 
\begin{equation}\label{eq:Egl:GFDT}
	\langle f_t^{\vphantom{\trans}} f_{t'}^\trans\rangle 
		= \sK_{t-t'}\langle x_0^{\vphantom{\trans}} x_0^\trans\rangle,
\end{equation}
where $t\geq t'$. 
Here, angular brackets indicate the average over an equilibrium ensemble of initial conditions, which, for the left hand side, is equivalent to the average over realizations of the stochastic force
\cite{zwanzig2001}. 
A central object of interest is the normalized equilibrium correlation matrix 
\begin{equation}
	\sZ_t = \langle x_t^{\vphantom{\trans}} x_0^\trans\rangle [\langle x_0^{\vphantom{\trans}} x_0^\trans\rangle]^{-1}. 
\end{equation}
Since $\langle f_t^{\vphantom{\trans}}x_0^\trans\rangle=0$, this quantity satisfies the homogeneous equation of motion 
\begin{equation}\label{eq:Egl:PropEq}
	\dot{\sZ}_t = \sV \sZ_ t + \tint \sK_{t'}\sZ_{t-t'}
\end{equation} 
and thus is identical to the propagator of the original generalized Langevin equation. 
Once $\sZ_t$ had been determined, the general solution of this equation can be obtained from the variation of parameters formula \eqref{eq:Fex:VarPara}.

\subsubsection{Example}\label{sec:EglExm}

The above formalism is particularly useful to incorporate memory effects into diffusion models \cite{porra1996,plyukhin2011,ishikawa2018}. 
As a concrete example, we here consider a Brownian particle in a one-dimensional harmonic potential with strength $\omega$.  
The relevant observables of this system are the position $\rx_t$ and the velocity $\rv_t$ of the particle, 
which we collect in the state vector $x_t = [\rx_t, \rv_t]^\trans$. 
The corresponding generalized Langevin equation reads 
\begin{equation}\label{eq:Egl:ExGLangevin}
	\dot{x}_t  = \sV x_t + \tint \sK_{t'} x_{t-t'} + f_t,
\end{equation}
with 
\begin{equation}
	\sV =   \omega
			\left[
				\begin{array}{rr}
 					0 & 1 \\ -1 & 0
				\end{array}
			\right],
			\quad
 	\sK_t =-\Gamma_t
 			\left[
 				\begin{array}{rr}
 					0 & 0 \\ 0 & 1
 				\end{array}
 			\right]
\end{equation}
and $f_t =[0,\xi_t]^\trans$, where we have chosen natural units for position and velocity so that $\langle \rx_0^2\rangle = \langle \rv_0^2\rangle =1$.  
The friction kernel $\Gamma_t$ and the stochastic force $\xi_t$ describe the influence of the fluid molecules surrounding the particle. 
The generalized fluctuation-dissipation theorem \eqref{eq:Egl:GFDT} now becomes 
\begin{equation}\label{eq:Egl:ExFDT}
	\langle \xi_t \xi_{t'} \rangle = \Gamma_{t-t'},
\end{equation}
where $t\geq t'$. 
For concreteness, we focus on the simple friction kernel
\begin{equation}
	\Gamma_t = M e^{-k t},
\end{equation}
where $M, k>0$ are free parameters. 
The propagator for the generalized Langevin equation \eqref{eq:Egl:ExGLangevin} can then be found explicitly by solving Eq.~\eqref{eq:Egl:PropEq} in Laplace space, which yields the result  
\begin{align}\label{eq:Egl:ExProp}
\sZ_t = \sum_{j=1}^3 \biggl(
 	\frac{k+\lb_j}{\Delta_j}
 		(\lambda_j + \sV)
 	 +	\frac{M}{\Delta_j}
 	 	\left[
 	 		\begin{array}{rr}
 				1 & 0 \\ 0 & 0 
 			\end{array}
 		\right]
 		\biggr) e^{\lb_j t}
\end{align}
with $\Delta_j = (\lb_j-\lb_{j+1})(\lb_j-\lb_{j+2})$ and $\lb_{3+1}=\lb_1$.
Here, the characteristic rates $\lb_j$ are given by the roots of the cubic equation 
\begin{equation}
	(\lb^2+\oa^2)(\lb+k) + \lb M =0. 
\end{equation}

\begin{figure}
\includegraphics[width=8.5cm]{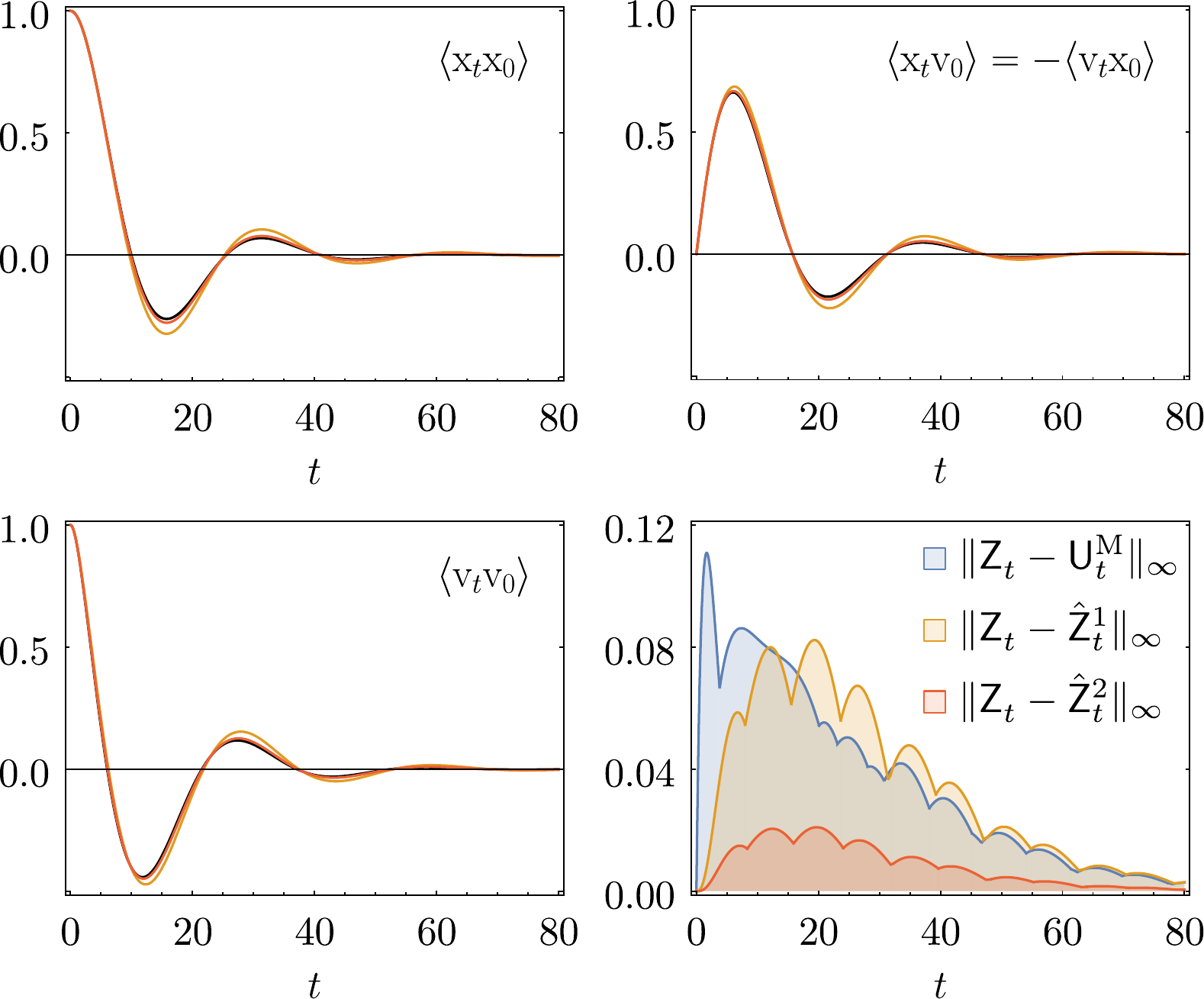}
\caption{\label{Fig:Egl}
Generalized Brownian motion in a harmonic potential.
For all plots, we have set $\oa=0.2$, $M=0.15$ and $k=1$ such that $\varepsilon=4M/(1-\omega)^2\simeq 0.94<1$ and all quantities are dimensionless.  
\textbf{Top left:} Position autocorrelation function. 
The black line indicates the exact solution \eqref{eq:Egl:ExProp}; the orange and red lines correspond to the first and second order approximations, \eqref{eq:Egl:FirstProp} and \eqref{eq:Egl:SecondProp}, respectively. 
\textbf{Top right:} Position-velocity cross-correlation function. 
\textbf{Bottom left:} Velocity autocorrelation function.
\textbf{Bottom right:} 
The orange and red curves indicate the error of the first- and second-order approximations of the propagator in the maximum absolute row sum norm. 
For comparison, the blue curve shows the error of the Markov approximation \eqref{eq:Egl:MarkovApprox}.
} 
\end{figure}

The exact result \eqref{eq:Egl:ExProp} provides a useful benchmark for the perturbation theory developed in Sec.~\ref{sec:Fpt}, which can be applied to the present model as follows. 
We choose $\nrm{\cdot}=\nrm{\cdot}_\infty$ to be the maximum row sum norm and set $k=1$ from here onward so that all relevant quantities become dimensionless. 
The weak-memory conditions \eqref{eq:Ftt:parabounds} then read $\oa<1$ and $\varepsilon= 4M/(1-\oa)^2 <1$. 
If these conditions are met, the successive approximations \eqref{eq:Fbr:EFExp} of the memory function $\sE_t$ converge uniformly over $\bR^+$.
Using the formulas \eqref{eq:Fbr:EInc} and \eqref{eq:Fbr:ERec}, we find that the first two terms of this sequence are given by 
\begin{align}
	\sE^1_t & = \sH_{11} e^{- t},\\
	\sE^2_t & = \sH_{21} e^{- t} + \sH_{22} t e^{-t}
\end{align}
with 
\begin{align}
	\sH_{11} & = \frac{M}{1 + \oa^2}
		\left[
			\begin{array}{rr}
				0 & 0 \\ \oa & 1
			\end{array}
		\right],\\
	\sH_{21} & = \sH_{11} + 
		\frac{M^2}{(1+\oa^2)^3}
			\left[
				\begin{array}{rr}
					0 & 0 \\2\oa & 1- \oa^2
				\end{array}
			\right],\\
	\sH_{22} & = \frac{M}{1+\oa^2}\sH_{11}.
\end{align}
Hence, the first and the second order approximations of the left generator are 
\begin{equation}
	\sL^1 = \sV - \sH_{11} 
		\quad\text{and}\quad
		\sL^2 = \sV - \sH_{21}
\end{equation}
according to Eq.~\eqref{eq:Fbr:GenApprox}. 
The corresponding approximations of the propagator, which can be obtained from the formulas \eqref{eq:Fbc:ZApprox} and \eqref{eq:Fbc:AApprox}, read
\begin{align}
	\label{eq:Egl:FirstProp}
 	\hat{\sZ}^1_t  = e^{\sL^1 t} &+ [\sL^1+\mathsf{1}]^{-1}(e^{(\sL^1 +1)t} + 1) \sH_{11} e^{-t},\\
 	\label{eq:Egl:SecondProp}
 	\hat{\sZ}^2_t  = e^{\sL^2 t} &+ [\sL^2+\mathsf{1}]^{-1}(e^{(\sL^2 +1)t} + 1) \sH_{21} e^{-t}\\
 			& + [\sL^2 + \mathsf{1}]^{-2}(e^{(\sL^2+\mathsf{1}) t} - (\sL^2 + 1) t -1)\sH_{22} e^{-t}.
 		\nonumber
\end{align}
In Fig.~\ref{Fig:Egl}, we compare these approximations with the exact propagator \eqref{eq:Egl:ExProp}. 
At short times, $t\lesssim 20$, both the first and the second-order approximations incur significantly smaller errors than the Markov approximation 
\begin{equation}\label{eq:Egl:MarkovApprox}
	\sU^\mathrm{M}_t = e^{\sL^1 t}.
\end{equation}
At long times, $t\gtrsim 50$, the errors of the first-order and Markov approximations become comparable, which is plausible, since both use the same generator. 
On any time scale, the second-order approximation is more accurate than both the first-order and Markov approximations by about a factor $4$, which is consistent with the value of the memory parameter being $\varphi = M \sim 1/6$. 

\section{Discussion and Perspectives}\label{sec:P}

\subsection{Main Objective}\label{sec:Pmo}

From a conceptual perspective, the central aim of this article is to provide a rigorous mathematical framework to describe the transition from non-local time evolution equations of the form
\begin{equation}\label{eq:Pbs:nleeq}
	\dot{x}_t = \sV x_t + \tint \sK_{t'} x_{t-t'},
\end{equation}
which naturally arise when inaccessible degrees of freedom are eliminated from microscopic models, to local time evolution equations of the type 
\begin{equation}\label{eq:Pbs:leeq}
	\dot{y}_t = \sL y_t.
\end{equation}
In the following, we very briefly summarize the main technical results that lie at the heart of this framework, before discussing some connections between our theory and the recent literature and pointing out selected challenges for future research.

\subsection{Main Results}\label{Pmr}

Theorem~\ref{thr:LG} contains two of our main results. 
First, the conditions \eqref{eq:Ftt:kernelbound} and \eqref{eq:Ftt:parabounds} on the memory kernel $\sK_t$ and the adiabatic generator $\sV$ provide a precise definition of a weak-memory regime, where a local generator $\sL$ and a non-singular slippage matrix $\sD$ exist such that any solution of Eq.~\eqref{eq:Pbs:nleeq} is arbitrary well approximated by the solution $y_t = e^{\sL t}\sD x_0$ of Eq.~\eqref{eq:Pbs:leeq} at sufficiently long times. 
Second, the error of this local approximation is subject to the universal bound \eqref{eq:Ftt:LTAErr}.  
In addition, we have shown that the solution of Eq.~\eqref{eq:Pbs:nleeq} satisfies a quasi-local time evolution equation of the form
\begin{equation}\label{eq:Pbs:quleeq}
	\dot{x}_t = \sL x_t + \sE_t x_0, 
\end{equation}
where, in the weak-memory regime, $\sL$ can be chosen such that the memory function $\sE_t$ decays exponentially in time. 
As our third main result, we find that the generator $\sL$ whose corresponding memory function $\sE_t$ satisfies the bound \eqref{eq:Fpa:EFBnd} is unique. 
Moreover, any other choice for the generator leads to a memory function that decays significantly slower than $\sE_t$, as the asymptotic relation \eqref{eq:Fpu:NonPropGen} shows. 
Finally, this optimal generator $\sL$, as well as its corresponding memory function $\sE_t$ and the slippage matrix $\sD=\sA_\infty=\sB_\infty$, can be constructed systematically through a convergent perturbation theory in the memory strength, where the Markov generator \eqref{eq:int:MarkovGen} is recovered in first order. 
This scheme, which is defined by the Eqs.~\eqref{eq:Fbr:EFExp}-\eqref{eq:Fbr:EFInc}, \eqref{eq:Fbr:GenApprox} and \eqref{eq:Fbc:ABApprox}, constitutes our fourth main result.

\subsection{Related Recent Results}\label{sec:Per}

Finding universal methods to convert, at least approximately, non-local time evolution equations of the form \eqref{eq:Pbs:nleeq} into local ones is a long-standing problem in the theory of open systems and a variety of techniques have been developed to this end
\cite{nestmann2021a,nestmann2021,bruch2021,contreras-pulido2012,karlewski2014,
schaller2008,schaller2009,majenz2013,farina2019,mozgunov2020,cattaneo2019,
nathan2020,davidovic2020,kleinherbers2020,trushechkin2021,davidovic2022}. 
A particularly elegant approach is the time-convolutionless, or TCL method, which seeks to construct a time dependent generator $\sL_t$ so that Eq.~\eqref{eq:Pbs:nleeq} is equivalent to the ordinary differential equation \cite{tokuyama1975,tokuyama1976,grabert1977,
grabert1978,chaturvedi1979,shibata1980,shibata1977,chaturvedi1979,shibata1980,breuer2001}
\begin{equation}
	\dot{x}_t = \sL_t x_t. 
\end{equation}
The relationship between the adiabatic generator $\sV$, the memory kernel $\sK_t$ and the TCL generator $\sL_t$ has been thoroughly investigated in the literature, see for instance Refs.~\cite{shibata1977,chaturvedi1979,shibata1980,breuer2001} and the more recent Refs.~\cite{vacchini2010,smirne2010,timm2011,megier2020}.
A very recent set of studies in this context is particularly relevant to our work \cite{nestmann2021a,nestmann2021,bruch2021}. 
The approach developed there rests on a time dependent fixed point equation for the TCL generator, which has a similar structure as Eq.~\eqref{eq:Fpg:LAsmp} and reduces to the latter in the long-time limit, provided that $\sL_t$ converges to a local generator $\sL$. 
This fixed point equation implies that the eigenvalues of $\sL$ have to satisfy the condition \eqref{eq:Fnp:EV} and leads to an explicit formula for the slippage matrix $\sD$ in terms of its spectral decomposition, which is equivalent to the expression given in Eq.~\eqref{eq:Fpg:DDef} if $\sL$ is diagonalizable with non-degenerate eigenvalues. 
Furthermore, the approach of Refs.~\cite{nestmann2021a,nestmann2021,bruch2021} makes it possible to derive a systematic perturbation theory for $\sL_t$ and $\sL$.  
The resulting series representation of $\sL$ can be regarded as a formal reordering of the memory expansions derived in Refs.~\cite{contreras-pulido2012,karlewski2014}, which, if they converge, sum to a solution of the fixed-point equation \eqref{eq:Fpg:LAsmp}, as has been shown in Ref.~\cite{nestmann2021a}. 
However, none of these studies systematically addresses questions of convergence and uniqueness. 
The present article contributes towards filling this gap as follows. 
First, we have shown that the weak-memory conditions \eqref{eq:Ftt:kernelbound} and \eqref{eq:Ftt:parabounds} are sufficient for the TCL generator to approach a time independent generator in the long-time limit. 
Second, this generator corresponds to the unique and attractive solution of the fixed-point equation \eqref{eq:Fpg:LAsmp} in the set $B_\rho$, which is defined in Eq.~\eqref{eq:Fpg:defBrho}. 
It is thus tempting to speculate that the conditions \eqref{eq:Ftt:kernelbound} and \eqref{eq:Ftt:parabounds} are also sufficient for the TCL generator to be a unique attractive solution of the time dependent fixed-point equation of Ref.~\cite{nestmann2021a}.
We leave it to future research to further investigate this conjecture. 

\subsection{Future Challenges}\label{sec:Pfc}
This article has primarily focused on the classical realm, where it opens various directions for future research. 
For instance, it would be interesting to analyze how accurately the presented approximation methods can reproduce the memory profile of the original non-local dynamics; 
this concept was recently developed in Refs.~\cite{lapolla2021,vollmar2024} to provide a model-free measure of memory effects. 
At the same time, however, our framework is readily applicable also to open quantum systems. 
Here, the role of the state vector is played by the density matrix of the system of interest and Eq.~\eqref{eq:Pbs:nleeq} as well as its local equivalents and approximations are collectively referred to as quantum master equations \cite{breuer2002}.  
A key challenge in this context is to find local generators that, besides leading to accurate approximations of the actual dynamics, gives rise to completely positive dynamical maps, i.e., propagators \cite{breuer2002}. 
This task is complicated by the fact that the memory kernel is typically not known explicitly, but instead must be obtained through perturbative expansions, whose convergence properties are generally not well understood. 
While, there exists a variety of advanced techniques that produce complete positivity preserving local generators, including, among many others, temporal coarse graining methods \cite{schaller2008,schaller2009,majenz2013,farina2019,mozgunov2020} and sophisticated improvements of the Markov and secular approximations \cite{cattaneo2019,nathan2020,davidovic2020,kleinherbers2020,trushechkin2021,davidovic2022}, most of these methods have essentially the same level of accuracy as the Markov approximation. 
In principle, our theory makes it possible to iteratively derive more accurate approximations through the scheme developed in Sec.~\ref{sec:Fpt}, at least if the memory kernel is known to sufficiently high order to ensure consistency of the perturbation theory.
At this point, there is no reason to expect that the resulting propagators \eqref{eq:Fbc:ZApprox} or their long-time versions, which appear in Eq.~\eqref{eq:Fbc:LTAApprox}, should in general correspond to completely positive maps. 
However, it appears plausible that, at any level of the perturbation theory, violations of positivity could be consistently of higher order in the memory strength. 
Developing this hypothesis into a rigorous statement presents an important problem for future research, as progress in this direction could help unify the existing range of quantum master equations and thus enhance our overall understanding of the dynamics of open quantum systems.  

A key limitation of our framework so far is its restriction to autonomous non-local time evolution equations, which precludes its application to systems that are subject to time dependent driving fields. 
For this class of systems, both the adiabatic generator and the memory kernel will typically acquire an explicit time dependence such that  Eq.~\eqref{eq:Pbs:nleeq} becomes \cite{fick1990,vrugt2020,schilling2022}
\begin{equation}\label{eq:Pbs:nleeqTD}
 	\dot{x}_t = \sV_t x_t + \tint \sK_{t,t'} x_{t-t'}. 
\end{equation}
It may still be possible to find a time dependent local generator $\sL_t$ so that Eq.~\eqref{eq:Pbs:nleeqTD} is equivalent to the quasi-local equation 
\begin{equation}
	\dot{x}_t = \sL_t x_t + \sE_t x_0.
\end{equation}
However, it is now no longer clear which conditions should be applied to define $\sL_t$, which properties can be expected from the memory function $\sE_t$ and whether the resulting scheme would still provide an advantage over the TCL approach, which already covers equations of the form \eqref{eq:Pbs:nleeqTD}. 
Similarly, we have exclusively focused on linear time evolution equations.
Generalized Langevin equations, for example, can however easily become non-linear in the variables of interest \cite{zwanzig2001}, and it is not obvious whether and how our framework can be extended to such situations.
In both cases, it may be possible to derive formal equations of motion for the memory function, which can be solved perturbatively in the memory strength. 
Nonetheless, it would then still be necessary to find appropriate boundary conditions for the individual terms of the expansion and to establish its convergence under suitable conditions. 
Further investigations in these directions can be expected to be challenging, but could eventually lead to a fully universal mathematical framework for the description of weak memory effects in small-scale systems.
Such a unifying theory may then incorporate existing methods for non-linear systems, like the correlation time expansion developed in Ref.~\cite{fox1983}.

Finally, our analysis has focused only on dynamical aspects, while thermodynamic consistency, that is, the possibility to equip a given time evolution equation with dynamical formulations of the first and the second law, is increasingly understood as an essential requirement for viable models of open systems \cite{esposito2012,bo2017,wachtel2018,strasberg2019,busiello2020,seiferth2020,potts2021,dann2021,soret2022,avanzini2023}.
In fact, the concepts of stochastic and quantum thermodynamics, which include fluctuation theorems, thermodynamic uncertainty relations and trade-off relations between the figures of merit of small-scale thermal machines, have transformed our understanding of micro and nanoscale systems over the past two decades \cite{strasberg2022,shiraishi2023}. 
It is therefore an appealing open question whether the quasi-local equation \eqref{eq:Pbs:leeq} can actually be equipped with a consistent thermodynamic structure if the generator and the memory function are constructed as in Sec.~\ref{sec:Fpt}.
If so, our theory could shed new light on the role of memory effects in small-scale thermodynamic processes, at least for autonomous systems, which can still be driven far away from equilibrium by thermal or chemical gradients.  

We conclude this article by stressing that the above list of future challenges is not meant to be complete, but rather to illustrate how our results provide starting points for investigations in various different directions. 
The central achievement of our work is to deliver a rigorous mathematical basis for the systematic treatment of weak but significant memory effects in a large class of micro and nanoscale systems. 
Further developing these ideas could well open new perspectives in the theory of open systems, both classical and quantum, and potentially enable wide-ranging unifications of existing theoretical methods. 

\begin{acknowledgments}
The author gratefully acknowledges insightful discussions with Paul Nemec, who verified the mathematical results reported in this article and corrected an error in the initial version of the part concerned with the uniqueness of the proper local generators.
This work was supported by the Medical Research Council (Grants No. MR/S034714/1 and MR/Y003845/1) and the Engineering and Physical Sciences Research Council (Grant No. EP/V031201/1).
The author further acknowledges support from the University of Nottingham through a Nottingham Research Fellowship. 
\end{acknowledgments}

\end{document}